\documentclass[12pt]{article}

\usepackage[margin=1in]{geometry}
\usepackage{amsmath}
\usepackage{amssymb}
\usepackage{graphicx}
\usepackage{placeins} 
\usepackage{booktabs}
\usepackage{subcaption}
\usepackage{xcolor} 
\usepackage[style=authoryear, backend=biber, natbib=true, giveninits=true, uniquename=false, uniquelist=false]{biblatex}
\usepackage[hidelinks]{hyperref}
\usepackage[nameinlink]{cleveref}   
\crefname{figure}{Figure}{Figures}      \Crefname{figure}{Figure}{Figures}
\crefname{table}{Table}{Tables}         \Crefname{table}{Table}{Tables}
\crefname{section}{Section}{Sections}   \Crefname{section}{Section}{Sections}
\crefname{subsection}{Section}{Sections} \Crefname{subsection}{Section}{Sections}
\crefname{appendix}{Appendix}{Appendices} \Crefname{appendix}{Appendix}{Appendices}
\crefname{equation}{equation}{equations} \Crefname{equation}{Equation}{Equations}
\usepackage{aastex_macros}
\usepackage{tikz}
\usetikzlibrary{decorations.pathmorphing, decorations.markings, arrows.meta, calc}

\usepackage{authblk}

\begin{document}
\title{Properties and limitations of linear models for mutual coupling in 21 cm arrays}
\author[1]{V. MacKay}
\author[2,3]{R. Pascua}
\author[1,4]{E. Rath}
\author[1,4]{J. N. Hewitt}
\author[5]{T. Cox}
\author[6]{N. Kern}
\author[5]{J. S. Dillon}
\author[5]{A. Parsons}
\author[7,8]{E. de Lera Acedo}
\author[7]{N. Fagnoni}
\affil[1]{Kavli Institute, Massachusetts Institute of Technology, Cambridge, MA, USA}
\affil[2]{Dunlap Institute, University of Toronto, Toronto, ON, Canada}
\affil[3]{Perimeter Institute for Theoretical Physics, Waterloo, ON, Canada}
\affil[4]{Department of Physics, Massachusetts Institute of Technology, Cambridge, MA, USA}
\affil[5]{Department of Astronomy, University of California, Berkeley, CA, USA}
\affil[6]{Department of Physics, University of Michigan, Ann Arbor, MI, USA}
\affil[7]{Cavendish Astrophysics, University of Cambridge, Cambridge, UK}
\affil[8]{Kavli Institute for Cosmology, University of Cambridge, Cambridge, UK}
\date{\today}

\maketitle

\begin{abstract}
Mutual coupling between elements of low-frequency arrays is a leading systematic for 21~cm cosmology, leaking bright foregrounds into otherwise clean regions of the power spectrum. Many mitigation models describe coupling as a linear combination of uncoupled signals re-radiated from neighboring feeds, but data suggest additional mechanisms. Inspired by antenna scattering theory, we decompose coupling into an \emph{antenna mode}, constructed from translated isolated responses and expected to be removable by linear combination models, and a \emph{structural mode}, the remainder. Full-wave simulations of HERA-like feeds and dishes in a 37-element hexagonal array show that the antenna mode accounts for only about $1\%$ of the beam perturbation power, suggesting limited scope for mitigation with linear models. Two effects nevertheless lead to substantially better suppression. First, correlating beams and weighting them by the sky changes the modes' relative importance: correcting the antenna mode can remove more than half of the coupling power in simulated visibilities. Second, relaxing the coefficient relationships imposed by re-radiation allows linear corrections to remove structural-mode coupling too. Beam-derived corrections can reduce coupling power by over an order of magnitude in our noiseless simulations of perfectly redundant arrays. Fitting directly to known isolated visibilities reduces residuals further on the fitted field, but this target is generally unavailable in observations. Practical correction therefore requires additional constraints to separate the unknown sky from coupling while preserving the cosmological signal. We introduce an extension of holographic back-projection to coupled beams, providing an intuitive picture of these perturbations in the aperture plane.
\end{abstract}

\section{Introduction}
\label{sec:introduction}

Large radio interferometric arrays are targeting the cosmological 21~cm signal across a wide range of redshifts, both to probe the Cosmic Dawn and Epoch of Reionization (EoR), and to map large-scale structure in the post-reionization Universe through to the present day. These experiments must contend with instrumental systematics whose control requirements are extraordinarily stringent: foregrounds exceed the signal by many orders of magnitude, and although they are intrinsically spectrally smooth, they acquire chromaticity through uncalibrated instrumental response, leaking power into the modes used to measure the cosmological signal. The field and its challenges are reviewed in \citet{furlanetto_et_al_2006, mesinger_2016, liu_shaw_2020}. While significant progress has been made---with increasingly sensitive upper limits at high redshift \citep{eastwood_et_al_2019, hera_collaboration_et_al_2023, mertens_et_al_2025, ceccotti_et_al_2025, nunhokee_et_al_2025, munshi_et_al_2025, hera_collaboration_et_al_2026}, and post-reionization detections in cross-correlation with galaxy surveys \citep{chime_collaboration_et_al_2023, cunnington_et_al_2023, chime_collaboration_et_al_2024} and in autocorrelation \citep{paul_et_al_2026, chime_collaboration_et_al_2025}---these instruments are often systematics- rather than noise-limited, especially on the largest scales.

Among the most challenging systematics is mutual coupling: electromagnetic interactions between antennas that modify their effective radiation patterns, a problem extensively studied in the antenna engineering literature \citep[reviewed by][]{craeye_gonzalez_ovejero_2011}. In a densely packed array, the resulting direction- and frequency-dependent beam perturbations introduce spectral structure into the measured visibilities that can overwhelm the cosmological signal with foregrounds over a broad range of spatial scales. Full-wave simulations of the HERA Phase~I receiver system likewise show that mutual coupling extends the instrumental delay response, with its extent depending on array size and antenna position \citep{fagnoni_et_al_2021b}. PAPER \citep{parsons_et_al_2010} and LOFAR \citep{van_haarlem_et_al_2013} already mention it as a possible limiting systematic, and subsequent 21~cm power-spectrum analyses have repeatedly named it as a leading candidate or principal source of excess power \citep{ali_et_al_2015, beardsley_et_al_2016, ewall_wice_et_al_2016, mertens_et_al_2020, hera_collaboration_et_al_2023, hera_collaboration_et_al_2026}.

Several mitigation strategies have been explored. A natural approach is to measure or simulate the coupled per-element responses and adopt them, in place of the idealized isolated response, in calibration and analysis. For the Murchison Widefield Array (MWA), embedded-element beam models account for coupling among the 16 dipoles within each tile \citep{sutinjo_et_al_2015,sokolowski_et_al_2017,nunhokee_et_al_2025}, and \citet{chokshi_et_al_2024} showed in simulations that foreground leakage from tile-to-tile beam variations can be mitigated if a precise model of each beam's perturbation is known and used in calibration. CHIME incorporates empirical beam models in its analysis, inferred from observations of catalogued radio sources and validated against independent holographic and solar measurements \citep{chime_collaboration_et_al_2023,chime_collaboration_et_al_2025}. LOFAR-EoR analyses fit direction-dependent station gains from the observations, accounting for beam and propagation effects when predicting and subtracting foreground emission \citep{gan_et_al_2023,mertens_et_al_2025}. Beam mapping and full-wave simulations remain demanding, especially at the precision required for cosmological sensitivity \citep{ohara_et_al_2025}. Recent accelerated solvers have nevertheless reduced the computational cost for large irregular arrays \citep{gueuning_et_al_2025}, enabling calculations of the embedded-element patterns of the 320-element HERA core \citep{gueuning_et_al_2026}.

A more direct strategy is to exclude heavily contaminated baselines from the analysis altogether. In MWA Phase~II, entire baseline classes were discarded because of a persistent common-mode systematic, for which mutual coupling remains a possible, albeit unconfirmed explanation \citep{kolopanis_et_al_2023,elder_et_al_2025}. CHIME's cosmological analyses omit intra-cylinder baselines partly because of enhanced cross-talk---a related form of coupling in which receiver noise, rather than sky signals, is radiated between feeds \citep{chime_collaboration_et_al_2025}. Such cuts trade sensitivity and baseline coverage for reduced contamination.

Another method leverages the fact that coupling signatures occupy characteristic regions of fringe-rate and delay space. \citet{kern_et_al_2019} developed and validated in simulations a subtraction method that models coupling systematics using a singular-value decomposition in time-delay space, restricting the model to low fringe rates to limit cosmological signal loss; \citet{kern_et_al_2020} then demonstrated it on HERA Phase~I data, for slowly varying systematics whose physical origin remained uncertain. \citet{rath_et_al_2025} found that HERA's fiducial main-lobe fringe-rate filter reduces coupling power in simulations by roughly a factor of~100 overall---though suppression is less effective on baselines with short east--west projections, because the desired sky signal varies slowly and overlaps the near-zero-fringe-rate component of the coupling; preserving that signal therefore retains more contamination. Even with this reduction, they find that the residual coupling contamination remains well above the simulated 21~cm signal across much of the EoR window.

A complementary framework---which is the focus of this work---models coupling as a linear transformation of uncoupled signals. Concretely, antenna $i$'s coupled voltage $v_i^c$ is written as a linear combination of uncoupled voltages $v_j^{(0)}$,
\begin{equation}\label{eq:coupled_voltage_linear_model}
  v_i^{\rm c} = v_i^{(0)} + \sum_j \epsilon_{ij}v_j^{(0)},
\end{equation}
where the $\epsilon_{ij}$ factors are \emph{coupling coefficients}. The corresponding visibility model is
\begin{equation}
  \mathbf V_{\rm cpl}
    = \mathbf M\,\mathbf V_{\rm iso}\,\mathbf M^{\dagger},
    \label{eq:visibility_mixing_matrix}
\end{equation}
where $N_{\rm ant}$ is the number of antennas and all matrices involved have size $N_{\rm ant}\times N_{\rm ant}$; the coupling matrix is $\mathbf M=\mathbf I+\boldsymbol\epsilon$, with $\mathbf I$ the identity and $\boldsymbol\epsilon$ a matrix populated with the coupling coefficients. The correction can thus be obtained either by determining $\mathbf M$ and inverting it, provided it is invertible, or by fitting entries of $\mathbf{M}^{-1}$. A more general form, which we identify with a horizontal bar over the symbols, consists in defining a much larger $N_{\rm bl}\times N_{\rm bl}$ matrix $\mathbf{\bar{M}}$ acting as
\begin{equation}
    {\mathbf{\bar{V}}}_{\rm cpl} = \mathbf{\bar{M}}\,\mathbf{\bar{V}}_{\rm iso},
    \label{eq:baseline_visibility_mixing_matrix}
\end{equation}
where here, $\bar{\mathbf V}$ collects the visibilities into an $N_{\rm bl}$-long vector. While it is always possible to write an $\mathbf{M}$ coupling matrix from \cref{eq:visibility_mixing_matrix} in \cref{eq:baseline_visibility_mixing_matrix}'s $\mathbf{\bar{M}}$-form via $\bar{M}_{(ij),(kl)} = M_{ik}M_{jl}^*$, the converse is not necessarily true; $\mathbf{\bar{M}}$ may not admit an antenna-level representation of the form in \cref{eq:visibility_mixing_matrix}. \Cref{eq:baseline_visibility_mixing_matrix} thus has more freedom to describe coupling, but is less constrained. Here $N_{\rm bl}$ counts the baseline entries used, including both orientations and any retained autocorrelations; nominally redundant baselines can be combined to reduce the matrix size, although this discards information about any differences in their responses that may be unaccounted for. The correction can be obtained by fitting an operator that acts as $\mathbf{\bar{M}}^{-1}$, without first determining $\mathbf{\bar{M}}$; if the model is imperfect or $\mathbf{\bar{M}}$ is singular, this fitted operator need not be an exact inverse.

\citet{kern_et_al_2019} developed the first semi-analytic, first-order model for re-radiative mutual coupling based on a 2-element model for HERA Phase I (equivalent to isolating elements from the $\mathbf{M}$ matrix to form $2\times2$ pairwise coupling matrices), showing good phenomenological agreement with the Phase I data. \citet{josaitis_et_al_2022} derived a semi-analytic first-order model for populating a full, dense $\mathbf{M}$-like matrix based on impedance mismatch and re-radiation, and \citet{rath_et_al_2025} found that it reproduces characteristic features of HERA data qualitatively, but the data's perturbations remain significantly brighter, suggesting, among other possible causes, additional coupling mechanisms not captured by re-radiation alone. \citet{chime_collaboration_et_al_2022} also applies a re-radiation model and highlights its limitations. For SKA-Low, \citet{ohara_et_al_2026} fit an antenna-level correction matrix, analogous to $\mathbf{M}^{-1}$, by approximating isolated responses with linear combinations of coupled beams over a chosen field of view. They propagate this beam-derived correction to visibilities through a transformation of the form in \cref{eq:visibility_mixing_matrix}. Their simulations show accurate reconstruction within the fitted region, while residual sidelobe contamination limits recovery of the EoR window for a full sky.

The model of \cref{eq:coupled_voltage_linear_model} implicitly describes \emph{antenna-mode} scattering---re-radiation originating at the neighboring feeds' ports, which produces translated copies of the isolated response. In reality, a significant fraction of the coupling may be \emph{structural-mode} scattering, arising from electromagnetic scattering off the physical structure of the neighboring elements, whose pattern need not entirely lie in the span of translated isolated responses, and is thus unreachable by the model.

In this paper, we introduce a framework for analyzing and decomposing mutual coupling that tests how much coupling linear corrections can remove in simulations based on full-wave isolated and coupled beam models. We first present an aperture-plane approach that localizes each neighbor's coupling contribution to compact features on the aperture plane, making the distinction between the antenna and structural modes visually explicit. We then show how much of the perturbation is attributable to either mode in a HERA-like array element's beam, and evaluate coupling removal through both coupling-matrix inversion and freely weighted linear visibility corrections. The latter need not result from self-consistent products of antenna-level correction coefficients; we examine the additional coupling removal this freedom permits and its dependence on the sky. \Cref{sec:coupling_in_aperture_plane} introduces the aperture-plane representation. \Cref{sec:antenna_structural_decomposition} defines the two coupling modes and presents the fraction of either mode in our simulated beams. \Cref{sec:coupling_in_kernel_space} asks how much coupling-matrix inversion and free-weight fits can remove in an idealized scenario where the coupled and uncoupled beams are known perfectly, and explores how baseline coverage may improve on those results. \Cref{sec:coupling_in_simulated_visibilities} evaluates these corrections in simulated visibilities, and \cref{sec:discussion} discusses the requirements for applying them to real data. \Cref{sec:conclusion} is a conclusion. Throughout the paper, we use a simplified HERA-like antenna model described in \Cref{app:cst_setup}.

\section{Mutual coupling in the aperture plane}
\label{sec:coupling_in_aperture_plane}

Fourier back-projection of complex beam maps is established in radio holography, where aperture-plane images reveal reflector illumination, blockage, and surface errors \citep{bennett_et_al_1976, scott_ryle_1977}. We extend its use to mutual coupling by back-projecting the difference between coupled and isolated beams. This gives a spatial view of perturbations with an intuitive physical interpretation, complementary to existing analysis approaches that look at their effects on S-parameters, angular-space beams, visibilities, and power spectra (e.g., \citealt{rath_et_al_2025,ohara_et_al_2026}).

The far-field radiation pattern of an antenna and its aperture-plane field distribution are related by a two-dimensional Fourier transform. Let $\vec E(\theta, \phi)$ be the complex far-field voltage pattern---where $\theta$ and $\phi$ are the zenith and azimuthal angles, respectively. We perform an orthographic projection of the beam's front hemisphere using direction-cosine coordinates $(\ell, m) = (\sin\theta\cos\phi, \sin\theta\sin\phi)$, and write the aperture-plane field as
\begin{equation}
  \vec A(x, y) = \iint_{\rm visible}\! \vec E(\ell, m) e^{-2\pi i (\ell x + m y)/\lambda} d\ell dm, 
  \label{eq:beam_to_aperture_transform}
\end{equation}
where $(x,y)$ are physical coordinates in the aperture plane and $\lambda$ is the free-space observing wavelength. With the normalization of \cref{eq:beam_to_aperture_transform}, the inverse relation is $\vec E(\ell,m)=\lambda^{-2}\iint\vec A(x,y) e^{2\pi i(\ell x+my)/\lambda}\,dx\,dy$. We refer to $\vec A$ as the \emph{reconstructed aperture-plane field}: it is the pattern that, when Fourier-transformed, reproduces the measured or simulated front hemisphere's far-field beam, and not a direct measurement of the physical fields on any real surface. The field $\vec A = (A_\theta, A_\phi)$ carries the two complex transverse components inherited from $\vec E$; wherever magnitudes appear we write $|\vec{A}|^2 \equiv |\vec{A}_\theta|^2 + |\vec{A}_\phi|^2$ for the pointwise power and $\Vert \vec{A}\Vert^2 \equiv \iint |\vec{A}|^2\, dx\, dy$ for its integral over the aperture plane. Inner products and least-squares fits likewise sum over the two components. For a single dish, $|\vec{A}|^2$ is mostly contained within a region of diameter $D$ (the physical dish diameter). Coupling to a neighbor at baseline vector $\vec{b}$ produces a perturbation to the aperture-plane field near $\vec{b}$ with spatial extent comparable to $D$.

\begin{figure*}[h!]
  \centering
  \includegraphics[width=\textwidth]{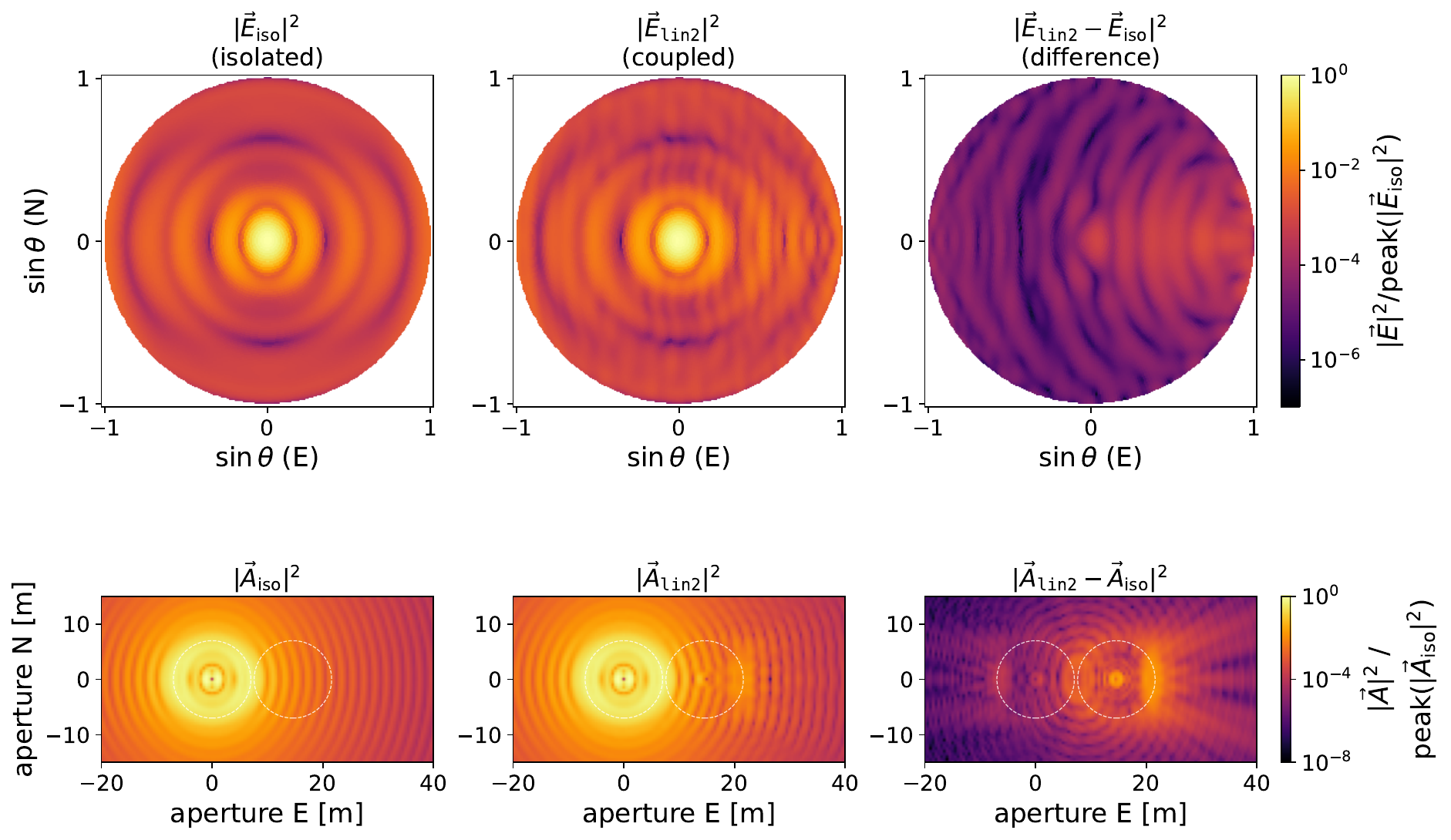}
  \caption{Orthographic projection (top row) and aperture-plane representation (bottom row) of the beam front hemisphere at 150~MHz. Isolated (first column) and two-element (second column) models, and their complex difference (third column). White dashed circles in the aperture row indicate the dish sizes and positions. The coupling perturbation appears as a broad fringe pattern in the orthographic view but as a localized feature in the aperture view. The (Gibbs) rings in the bottom row result from Fourier transforming the finite-support orthographic projection, and do not represent real physical features.\label{fig:beam_and_aperture_coupling_comparison}}
\end{figure*}

To illustrate this approach, we first show two simulations using a simplified HERA element model \citep{fagnoni_et_al_2021}, which consists of a Vivaldi-style feed floating above a 14~m dish---see \cref{app:cst_setup} for simplification details and model images. In all simulations, we only excite the feed plane aligned with the north--south axis. One simulation contains a single isolated element (labeled ``iso''), while the second includes two (\texttt{lin2}), separated by 14.6~m, based on HERA's shortest east--west baseline. \Cref{fig:beam_and_aperture_coupling_comparison} illustrates the beam-to-aperture relationship for the two simulations, along with their complex difference, at 150~MHz. The orthographic view (top row) shows the broad fringe pattern induced by the coupled neighbor across the sky; the aperture view (bottom row) localizes the same perturbation to a compact, dish-sized feature near the neighbor's position at $(14.6, 0)$~m, mostly visible in the difference plot (third column). In all aperture plots, concentric Gibbs ringing appears as the result of Fourier transforming a function with finite support; apodizing the projection would attenuate that feature, at the expense of attenuating horizon sensitivity and biasing quantitative results based on those aperture planes (see, e.g., \cref{sec:coupling_in_kernel_space}); plus, it would smear out fine features in the aperture plane. We therefore use unwindowed transforms throughout.

\begin{figure*}[h!]
  \centering
  \includegraphics[width=\textwidth]{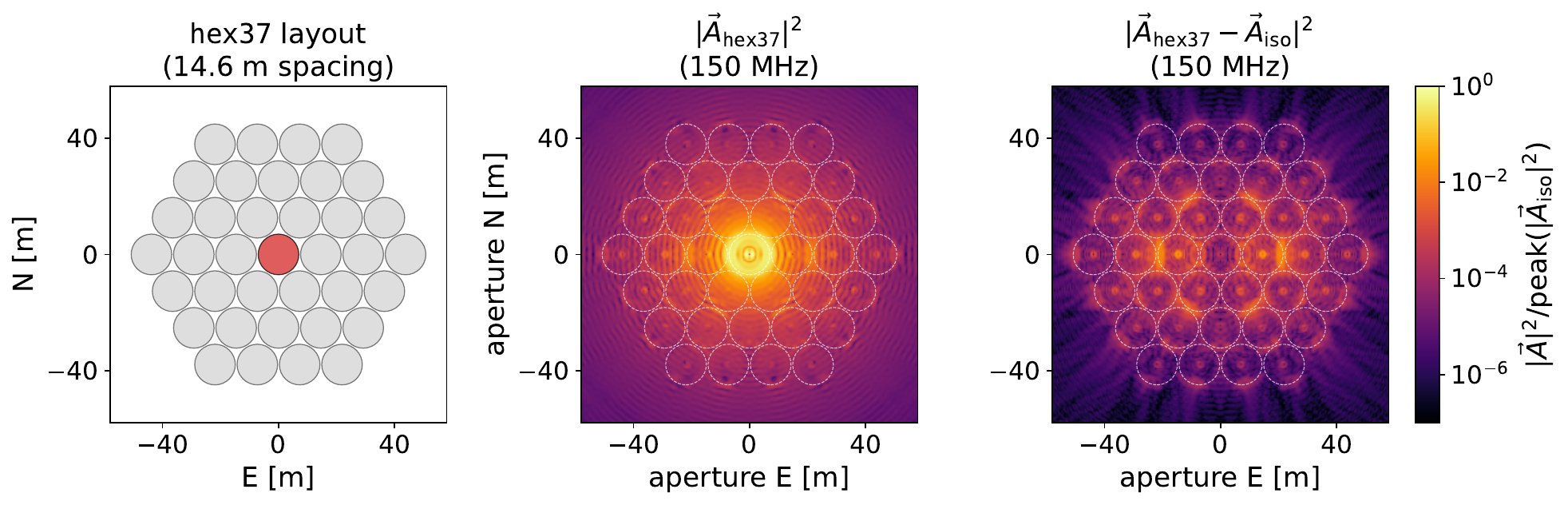}
 \caption{The \texttt{hex37} reference array. Left: the layout---37 elements in a close-packed arrangement with 14.6~m nearest-neighbor spacing; the highlighted central element is the one whose response is simulated. Center: the aperture power $|\vec A_{\texttt{hex37}}|^2$ of the coupled central element at 150~MHz. Right: the complex difference against the isolated element, $|\vec A_{\texttt{hex37}} - \vec A_{\rm iso}|^2$. Dashed circles mark the dish outlines.\label{fig:hex37_layout_and_coupled_aperture}} %
\end{figure*}

The feature in the bottom right of \cref{fig:beam_and_aperture_coupling_comparison} is our first diagnostic of the coupling's nature. If each neighbor simply re-radiated a copy of the signal it received, as assumed in \cref{eq:coupled_voltage_linear_model}, the aperture-plane difference would just be a faint, faithful copy of the isolated pattern (bottom left), shifted onto the neighbor's position---with possibly an even fainter copy at the origin, for higher-order reflections. Here, the perturbation carries an extra, asymmetric feature on the far rim of the neighbor's dish. The same representation scales from a pair to a full array. \Cref{fig:hex37_layout_and_coupled_aperture} shows our reference array---37 elements in a close-packed layout at 14.6~m shortest spacing, henceforth \texttt{hex37}---together with the aperture power of its coupled central element and the difference against the isolated element: one compact, dish-sized feature appears at each neighbor, each carrying that neighbor's contribution to the coupling.

We note that multiple reflections can introduce long propagation delays without placing coupling features at correspondingly greater distances from the origin in the aperture plane. Indeed, their locations reflect where the incident sky radiation first interacts with the array, whether at a feed or elsewhere on its structure; subsequent reflections change each contribution’s complex amplitude and frequency dependence without moving that point of entry. The coupling features therefore remain associated with the array’s physical structures (except for the ringing described above) rather than the accumulated propagation distance.

\section[Decomposition into antenna and structural modes]{Decomposition into antenna and structural modes}
\label{sec:antenna_structural_decomposition}

\subsection{Definitions}
\label{subsec:scattering_mode_definitions}

\begin{figure}[h!]
\centering
\begin{subfigure}[b]{0.82\linewidth}
\centering
\begin{tikzpicture}[
    dish/.style={thick, black},
    feed/.style={fill=black, circle, inner sep=2pt},
    sky arrow/.style={-{Stealth[length=3mm]}, blue!90!black, thick,
        decorate, decoration={snake, amplitude=0.5mm, segment length=4mm, post length=2mm}},
    direct arrow/.style={-{Stealth[length=3mm]}, green!80!black, thick,
        decorate, decoration={snake, amplitude=0.5mm, segment length=4mm, post length=2mm}},
    indirect arrow/.style={-{Stealth[length=3mm]}, red!90!black, thick,
        decorate, decoration={snake, amplitude=0.5mm, segment length=4mm, post length=2mm}},
    label style/.style={font=\small},
    scale=0.45
  ]
  \draw[dish] (-10.5,0) parabola bend (-3.5,-2.5) (3.5,0);
  \node[feed, label={[label style]above:feed $i$}] (feedi) at (-3.5,1.5) {};
  \draw[thick, gray] (-3.5,-2.5) -- (feedi);

  \draw[dish] (3.5+0.6,0) parabola bend (3.5+7.6,-2.5) (3.5+14.6,0);
  \node[feed, label={[label style]above:feed $j$}] (feedj) at (10.5,1.5) {};
  \draw[thick, gray] (10.5,-2.5) -- (feedj);

  \draw[sky arrow] (3.5+7.6+2,4) -- (3.5+7.6+2,-2.3);
  \draw[sky arrow] (3.5+7.6+2,-2.3) -- (10.7,1.3);
  \draw[sky arrow] (3.5+7.6+3,4) -- (3.5+7.6+3,-2.1);
  \draw[sky arrow] (3.5+7.6+3,-1.6) -- (10.7,1.3);
  \draw[sky arrow] (3.5+7.6+4,4) -- (3.5+7.6+4,-1.7);
  \draw[sky arrow] (3.5+7.6+4,-1.5) -- (10.7,1.3);
  \draw[blue!90!black, densely dashed, thin] (12.7,4.2) -- (18.5,4.2);
  \node[blue!90!black, label style, above] at (15.6,4.3) {incident sky wave};

  \draw[direct arrow] (10.2,1.7) -- (-3.2,1.7);
  \node[green!80!black, label style, above] at (3.5,2.1) {direct path};

  \draw[indirect arrow] (10.2,1.3) -- (3.5+0.3,0);
  \draw[indirect arrow] (3.5+0.3,0) -- (-3.2,1.3);
  \node[red!90!black, label style, above] at (3.5,0.4) {indirect path};

  \node[label style] at (-3.5,-3.5) {dish $i$};
  \node[label style] at (10.5,-3.5) {dish $j$};
\end{tikzpicture}
\caption{}
\label{subfig:antenna_mode_scattering_paths}
\end{subfigure}
\vspace{0em}
\begin{subfigure}[b]{0.82\linewidth}
\centering
\begin{tikzpicture}[
    dish/.style={thick, black},
    feed/.style={fill=black, circle, inner sep=2pt},
    sky arrow/.style={-{Stealth[length=3mm]}, blue!90!black, thick,
        decorate, decoration={snake, amplitude=0.5mm, segment length=4mm, post length=2mm}},
    scatter arrow/.style={-{Stealth[length=3mm]}, orange!80!black, thick,
        decorate, decoration={snake, amplitude=0.5mm, segment length=4mm, post length=2mm}},
    label style/.style={font=\small},
    scale=0.45
  ]
  \draw[dish] (-10.5,0) parabola bend (-3.5,-2.5) (3.5,0);
  \node[feed, label={[label style]above:feed $i$}] (feedi) at (-3.5,1.5) {};
  \draw[thick, gray] (-3.5,-2.5) -- (feedi);

  \draw[dish] (3.5+0.6,0) parabola bend (3.5+7.6,-2.5) (3.5+14+0.6,0);
  \node[feed, opacity=0.5, label={[label style, opacity=0.5]above:feed $j$}] (feedj) at (11.1,1.5) {};
  \draw[thick, gray] (11.1,-2.5) -- (feedj);

  \draw[blue!90!black, densely dashed, thin] (12.7,4.2) -- (18.5,4.2);
  \node[blue!90!black, label style, above] at (15.6,4.3) {incident sky wave};
  \draw[sky arrow] (18.1,4) -- (18.1,0);
  \begin{scope}[transparency group, opacity=0.5]
    \foreach \x/\y in {13.1/-2.296,14.1/-2.041,15.1/-1.684} {
      \draw[sky arrow] (\x,4) -- (\x,\y);
      \draw[sky arrow] (\x,\y) -- (11.3,1.3);
    }
  \end{scope}

  \fill[orange!80!black] (18.1,0) circle (2pt);
  \draw[scatter arrow, preaction={draw=white, line width=2.4pt, -}] (18.1,0) -- (-3.2,1.4);
  \node[orange!80!black, label style, above] at (3.5,1.5) {edge scattering};

  \node[label style] at (-3.5,-3.5) {dish $i$};
  \node[label style] at (10.5+0.6,-3.5) {dish $j$};
\end{tikzpicture}
\caption{}
\label{subfig:structural_mode_scattering_paths}
\end{subfigure}
\caption{Scattering paths for the two different coupling modes. (a) Antenna mode: a signal (blue) enters feed~$j$ and is partially reflected at the feed terminals. The re-radiated signal reaches feed~$i$ via direct (green) or indirect (red) paths, which we group together. This coupling mode is load dependent: altering the impedance at feed~$j$ modulates its complex amplitude. (b) Structural mode: the same incident sky wave is received at feed~$j$ and scatters off the far rim of dish~$j$ (orange) into feed~$i$, without passing through $j$'s feed port. The scattered contribution at $i$ can therefore correlate with the directly received signal at $j$. For the edge-scattered path shown, this contribution is independent of $j$'s feed termination.}
\label{fig:antenna_and_structural_scattering_paths}
\end{figure}
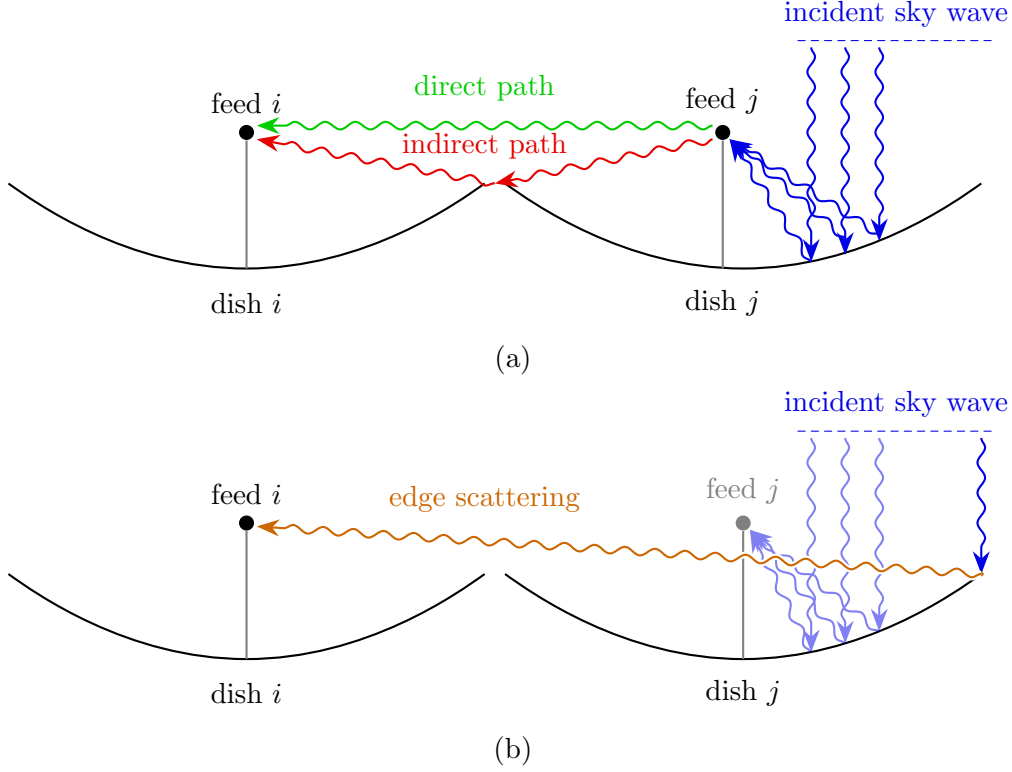

Mutual coupling can be modeled as an additive field superposed on the antenna's primary far-field beam. We describe this field as the result of scattering, and borrow from the antenna scattering literature \citep{green_1963, hansen_1989, collin_2003, knott_et_al_2004, best_kaanta_2009}, which separates scattered fields into two modes---antenna and structural. We reuse those labels to define two corresponding coupling modes. \emph{Antenna-mode} coupling occurs when a signal received at a neighboring feed reaches its terminals, and some fraction, set by the load impedance, is re-radiated into free space, ending at the dish under test via any path (\Cref{subfig:antenna_mode_scattering_paths}). This last qualifier is important: in this paper, we purposefully do not keep track of the number or type of reflections as a signal travels from one feed terminal to another, but instead group all possible paths together; this framework is detailed in \Cref{app:all_paths_coupling_coefficients}. We represent the antenna-mode scattered field as a load-dependent copy of the neighbor's isolated beam, so that the resulting coupling feature can be written as a linear combination of the neighboring elements' beams; \cref{eq:coupled_voltage_linear_model} assumes coupling of this mode. \emph{Structural-mode} coupling, by contrast, begins with scattering off the neighbor's physical structure, an interaction that does not require excitation of its feed port and would persist even if the structure were not functioning as an antenna at all (\Cref{subfig:structural_mode_scattering_paths}). It is defined as whatever remains of the total scattered field once the antenna-mode contribution is removed. 

In the single-neighbor case, one writes the scattered field as
\begin{equation}
  \begin{aligned}
    \vec E_{\rm scattered}(Z_L) &= \vec E_{\rm antenna}(Z_L) + \vec E_{\rm structural} \\
    &= \epsilon(Z_L)\,e^{2\pi i(b_x\ell + b_y m)/\lambda}\,\vec E_{\rm iso} + \vec E_{\rm structural},
  \end{aligned}
  \label{eq:load_dependent_scattered_field_decomposition}
\end{equation}
where $\vec b=(b_x,b_y)$ is the baseline vector from the element under consideration to its neighbor. The phase factor accounts for that displacement, so the antenna-mode term corresponds to a translated isolated response in the aperture plane, with the load dependence described by a complex scalar $\epsilon(Z_L)$. This scalar includes the phase accumulated along the coupling paths, but not the direction-dependent interferometric phase written explicitly here. We use $\epsilon$ to denote this scalar deliberately, as it is precisely the coupling coefficient that appeared in \cref{eq:coupled_voltage_linear_model}. In scattering literature, a reference load $Z_L'$ is chosen and the structural field is defined as $\vec{E}_{\rm structural}\equiv\vec{E}_{\rm scattered}(Z_L')$; the antenna-mode coefficient therefore satisfies $\epsilon(Z_L')=0$ by construction. Unless stated otherwise, all passive antenna ports in our simulations are terminated at a reference impedance $Z_0=100~\Omega$; the ``total'' (including both modes) coupled beam is the response with all neighbors at $Z_0$, so that $\epsilon(Z_0)$ is the antenna-mode coefficient at this operating load.

\subsection{A new definition of the antenna-mode coefficients}
\label{subsec:projected_antenna_mode_definition}

The scattering literature offers at least four conventions for the special load $Z_L'$, and hence for the antenna-mode coefficient and the decomposition itself; we review them and compare their results in \cref{app:antenna_mode_load_conventions}. However, rather than choosing among them, we use this ambiguity as license to define the decomposition by what we need it to do: make explicit how much of the coupling perturbation can be represented by a linear combination of isolated responses. This fifth convention, which we call the \emph{projected} definition, has an intuitive form in the aperture plane, where the antenna mode is obtained by projecting the total perturbation onto the isolated pattern translated to the neighbor's position. We first demonstrate the convention in the simple \texttt{lin2} system; in \cref{subsec:array_aperture_mode_decomposition} we generalize to larger arrays. We define 
\begin{equation}
  \epsilon_{\rm proj}(\nu) = \frac{\langle \vec A_{\rm iso}(\vec x - \vec b),\,\vec A_{\rm tot}(\vec x)-\vec A_{\rm iso}(\vec x)\rangle}{\|\vec A_{\rm iso}(\vec x)\|^2},
  \label{eq:single_neighbor_projection_coefficient}
\end{equation}
where $\vec b$ is the $14.6$~m east--west baseline vector for the \texttt{lin2} system. The inner product is defined over the aperture plane as $\langle \vec F,\vec G\rangle \equiv \sum_\beta\iint F_\beta^*(x,y)\,G_\beta(x,y)\,dx\,dy$, where $\beta\in\{\theta,\phi\}$ labels the E-field components. The structural mode is the residual orthogonal to this translated isolated-response template,
\begin{equation}
  \Delta \vec A_{\rm sm}(\vec x) \equiv \left(\vec A_{\rm tot}(\vec x) - \vec A_{\rm iso}(\vec x)\right) - \epsilon_{\rm proj}\, \vec A_{\rm iso}(\vec x - \vec b).
  \label{eq:projected_structural_aperture_residual}
\end{equation}
The decomposition satisfies
\begin{equation}
  \Delta \vec A = \Delta \vec A_{\rm am} + \Delta \vec A_{\rm sm}, \qquad \langle \Delta \vec A_{\rm am}, \Delta \vec A_{\rm sm} \rangle = 0.
  \label{eq:orthogonal_aperture_mode_decomposition}
\end{equation}
With this projected definition, the structural mode is precisely the perturbation left after removing everything that is shaped like a translated copy of the isolated response. We thus associate the structural mode with an aperture-space residual, the fraction of the original perturbation power that remains when the antenna mode is removed,
\begin{equation}
  r_\epsilon(\nu) \equiv
  \frac{\|\Delta\vec A-\Delta\vec A_{\rm am}\|^2}{\|\Delta\vec A\|^2}
  =\frac{\|\Delta\vec A_{\rm sm}\|^2}{\|\Delta\vec A\|^2},
  \label{eq:aperture_projection_residual_fraction}
\end{equation}
using the same aperture-plane norm as in the projection.

Like the definitions from the literature, the projected definition makes the antenna-mode field proportional to the translated isolated response, but additionally makes the structural residual orthogonal to that template by construction. The same fit could be performed in the far field, using the phase-shifted isolated response in \cref{eq:load_dependent_scattered_field_decomposition} and integrating over $d\ell\,dm$: by Parseval's theorem, the aperture-plane inner product and squared norm both acquire the same factor $\lambda^2$, which cancels in the fitted coefficient. We use the aperture-plane form because it makes the translated templates, and hence the content represented by the model, easier to visualize.

This fifth, ``projected'' definition is the most useful one for our purposes for two reasons. It defines a unique field for the specified inner product using only the isolated and multi-element simulations---without requiring a separate simulation with load $Z_L'$---and it absorbs into the antenna mode exactly the portion of the perturbation that is shaped like a translated isolated response, no more and no less. Meanwhile, its physical interpretation comes with two caveats: the projected antenna-mode field can pick up scattering that is structural and load independent in nature, but that happens to overlap with the span of the translated isolated pattern; and unlike the four load conventions of \Cref{tab:antenna_mode_reference_loads} (\cref{app:antenna_mode_load_conventions}), the projected coefficient has no closed-form expression in terms of the antenna and reference impedances---it is defined operationally, from the full-wave simulations. Neither caveat affects the mitigation strategies outlined in this paper.

\subsection{Decomposition in the aperture plane for arrays}
\label{subsec:array_aperture_mode_decomposition}

The single-neighbor projection of \cref{eq:single_neighbor_projection_coefficient} extends directly to a full array. For a central element $i$ in an array of $N$ antennas at positions $\{\vec p_j\}_{j=1}^{N}$ (taken relative to the central element, so that one of them, $\vec p_i = 0$, is the central element itself, included to capture self-coupling), we define the antenna-mode part of the perturbation as a complex-weighted sum of isolated patterns shifted to each antenna's position,
\begin{equation}
  \Delta \vec A_{\rm am}(\vec x) = \sum_{j=1}^{N} \epsilon_{j}\, \vec A_{\rm iso}(\vec x - \vec p_j),
  \label{eq:array_antenna_mode_template_expansion}
\end{equation}
with the coefficients $\boldsymbol{\epsilon}_{\rm proj} = (\epsilon_1,\dots,\epsilon_{N})^{\top} \in \mathbb{C}^{N}$ obtained at each frequency by jointly minimizing $r_\epsilon(\nu)$ in \cref{eq:aperture_projection_residual_fraction} over all $N$ complex coefficients. Since the denominator is independent of the coefficients, this is the least-squares fit of the total perturbation $\Delta \vec A = \vec A_{\rm tot} - \vec A_{\rm iso}$ onto the shifted-isolated basis. Explicitly, the coefficients solve the normal equations
\begin{equation}
  \sum_{k=1}^{N} G_{jk}\, \epsilon_k = c_j,
  \qquad
  G_{jk} = \bigl\langle \vec A_{\rm iso}(\vec x - \vec p_j),\, \vec A_{\rm iso}(\vec x - \vec p_k) \bigr\rangle,
  \qquad
  c_j = \bigl\langle \vec A_{\rm iso}(\vec x - \vec p_j),\, \Delta \vec A \bigr\rangle,
  \label{eq:array_projection_normal_equations}
\end{equation}
i.e.\ the projection of \cref{eq:single_neighbor_projection_coefficient} generalized to the (nearly diagonal) Gram matrix of the shifted templates. This multi-neighbor fit preserves the orthogonality between the structural mode and the span of the translated templates.

\begin{figure*}[h!]
  \centering
  \includegraphics[width=\textwidth]{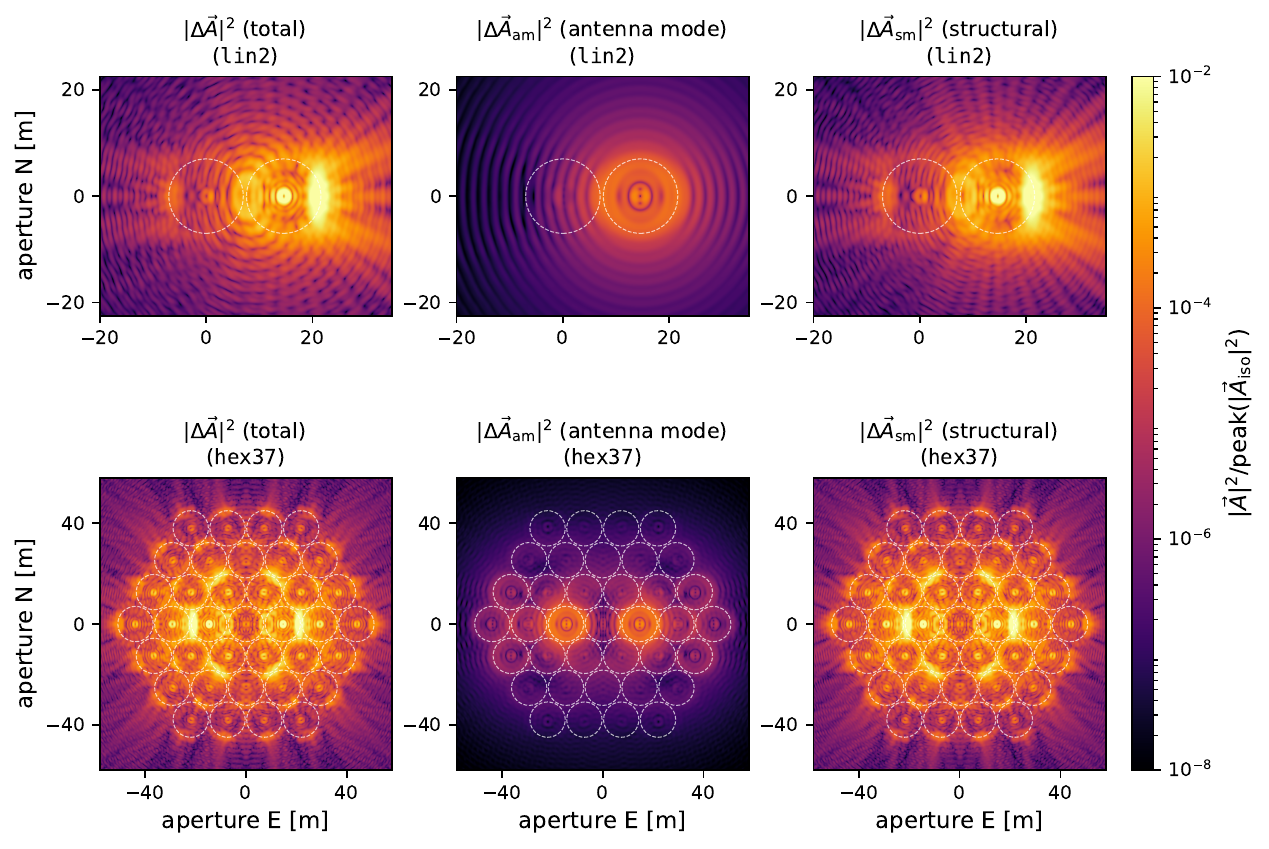}
  \caption{Projected decomposition of the coupling perturbation at 150~MHz, for \texttt{lin2} (top row) and \texttt{hex37} (bottom row), in the dish model. Columns: the total perturbation $|\Delta \vec A|^2$, its antenna-mode part $|\Delta \vec A_{\rm am}|^2$ (a superposition of translated copies of the isolated pattern, one per antenna position), and the structural-mode residual $|\Delta \vec A_{\rm sm}|^2$, normalized to the peak of the isolated-dish aperture power.\label{fig:dish_aperture_mode_decomposition}} %
\end{figure*}

\Cref{fig:dish_aperture_mode_decomposition} shows the resulting decomposition at 150~MHz, for \texttt{lin2} (top row) and \texttt{hex37} (bottom row): the total perturbation $|\Delta \vec A|^2$, its antenna-mode part $|\Delta \vec A_{\rm am}|^2$, and the structural-mode residual $|\Delta \vec A_{\rm sm}|^2$. In both systems the antenna-mode field is a superposition of translated isolated patterns, yet the individual copies appear slightly distorted in the \texttt{hex37} power map (lower center) because their Gibbs ringing patterns overlap and interfere. The structural mode carries almost all of the perturbation power and is concentrated toward the neighbors' rims, as well as near the dish centers, suggesting a contribution from HERA's large Vivaldi feeds. %

\begin{figure}[h!]
  \centering
  \includegraphics[width=\textwidth]{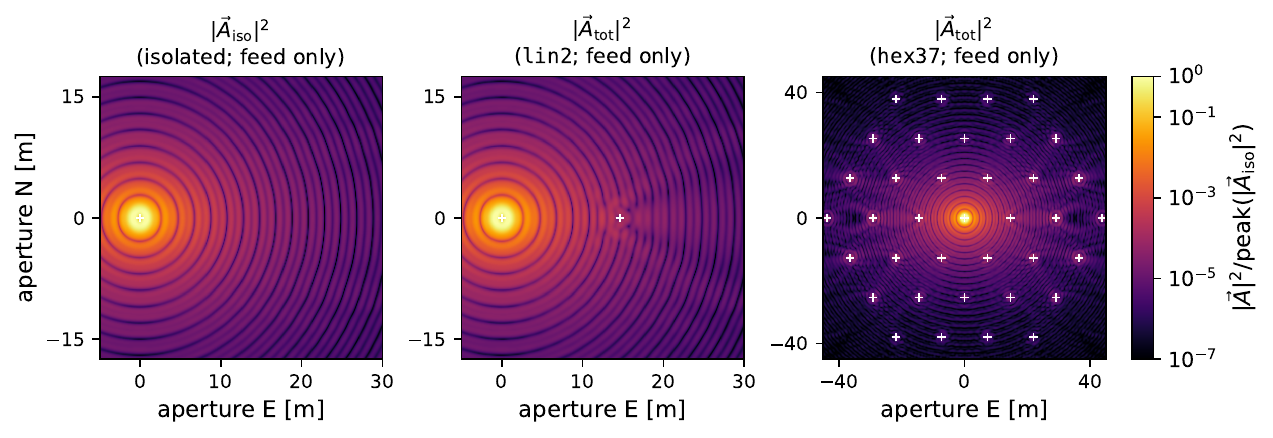}
  \caption{Aperture power at 150~MHz for the feed-only models. From left: the isolated feed, the coupled feed in \texttt{lin2}, and the coupled central feed in \texttt{hex37}. All three responses are normalized to the isolated-feed peak and share the same color scale; crosses mark the feed positions. The \texttt{hex37} panel covers a larger region of the aperture plane.}\label{fig:feed_only_aperture_responses}
\end{figure}

\begin{figure}[h!]
  \centering
  \includegraphics[width=\textwidth]{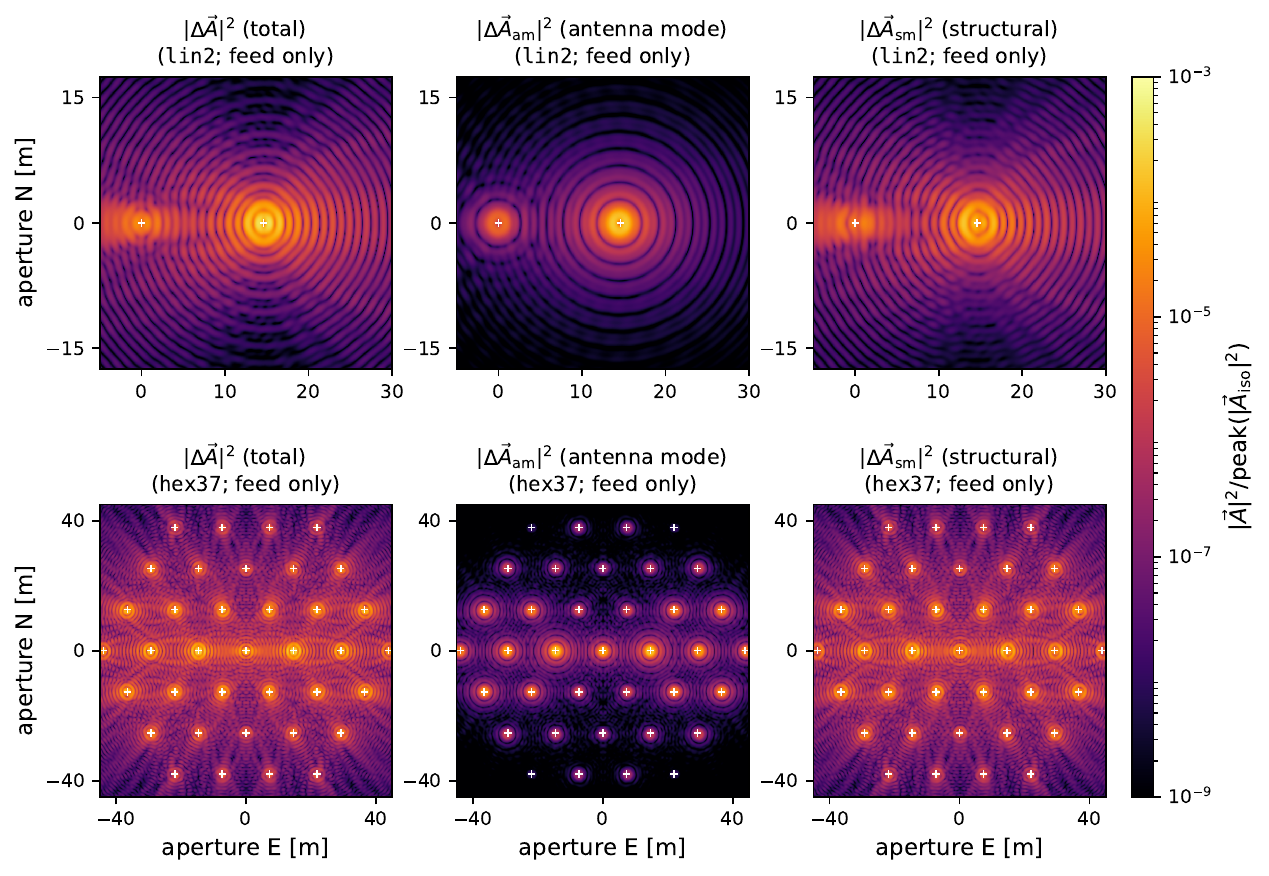}
  \caption{The projected decomposition of \cref{fig:dish_aperture_mode_decomposition}, repeated for the feed-only models at 150~MHz: \texttt{lin2} (top row) and \texttt{hex37} (bottom row). Columns: the total perturbation $|\Delta\vec A|^2$, its antenna-mode part $|\Delta\vec A_{\rm am}|^2$, and the structural-mode residual $|\Delta\vec A_{\rm sm}|^2$. All panels share the same color scale, normalized to the isolated-feed aperture-power peak; crosses mark the feed positions.}\label{fig:feed_only_aperture_mode_decomposition}
\end{figure}

To assess how the dishes and the feeds each contribute to structural coupling, we repeat the simulations with bare Vivaldi feeds pointing toward the sky in the same layouts (an isolated feed, \texttt{lin2}, and \texttt{hex37}). \Cref{fig:feed_only_aperture_responses} shows the isolated and coupled responses, and \cref{fig:feed_only_aperture_mode_decomposition} their mode decomposition. Compared with the simulations that include dishes, the feed-only maps show much brighter antenna-mode features relative to the structural residual, suggesting that the dishes contribute substantially to structural coupling. Removing the dishes does, however, change both the coupled response and the isolated reference, so this comparison does not directly measure a dish-only scattering contribution. A structural residual also remains in the feed-only simulations, which is plausible for a wideband Vivaldi feed with substantial conducting structure, including the perpendicular-polarization assembly and portions that may be active at one frequency but structural at another.

\begin{figure}[h!]
  \centering
  \includegraphics[width=\textwidth]{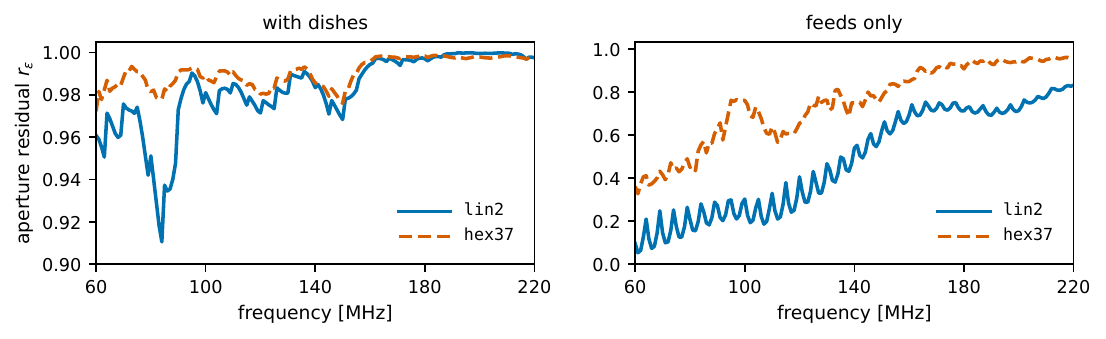}
  \caption{Aperture-space residual power fraction $r_\epsilon$ (\cref{eq:aperture_projection_residual_fraction}) after removing the projected antenna mode, for the dish (left) and feed-only (right) models. Solid curves show \texttt{lin2} and dashed curves show \texttt{hex37}. The dish panel is zoomed near unity.\label{fig:aperture_mode_power_budget}} %
\end{figure}

\Cref{fig:aperture_mode_power_budget} shows the aperture-space residual $r_\epsilon$ across the band for \texttt{lin2} and \texttt{hex37} (\cref{fig:hex37_layout_and_coupled_aperture}), in both the dish and feed-only models. Its complement, $1-r_\epsilon$, is the antenna-mode power fraction, with band medians of about $1.5\%$ for \texttt{lin2} and $0.8\%$ for \texttt{hex37} in the dish models; the corresponding feed-only values are $50\%$ and $24\%$. The higher antenna-mode fraction in the \texttt{lin2} layout is consistent with its single neighbor lying along the east--west direction preferred by the polarization analyzed here, whereas \texttt{hex37} includes neighbors at other azimuths, averaging out that effect.

\section[Coupling correction in uv space]{Coupling correction in $uv$ space}
\label{sec:coupling_in_kernel_space}

The previous sections examine coupling as a perturbation to each element's beam, but interferometric visibilities involve beam \emph{products}, so the aperture-plane budget does not by itself tell us how much contamination a linear model can remove. We now propagate the perturbations into those products and their Fourier transforms, the \emph{beam kernels}, to test the model before introducing sky convolution and baseline sampling in \cref{sec:coupling_in_simulated_visibilities}.

\subsection{Beam kernels}
\label{subsec:beam_kernel_definition}

The visibility measured by two elements with far-field voltage patterns $\vec E_i(\hat s)$ and $\vec E_j(\hat s)$ depends on their separation in wavelengths. We assume a coplanar array and choose the $z$ axis normal to its plane, so that $b_z=0$. We write $\vec u=(u,v)=(b_x,b_y)/\lambda$, where $\lambda$ is the observing wavelength; frequency dependence is omitted thereafter for simplicity. Restricting the sky integral to the front hemisphere, as in our beam reconstruction, gives
\begin{equation}
  V_{ij}(\vec u)= \int_{\rm front} T(\hat s)\,\vec E_i(\hat s)\cdot\vec E_j^*(\hat s)\,e^{-2\pi i(u\ell+vm)}\,d^2\hat s,
  \label{eq:visibility_sky_integral}
\end{equation}
where $\hat s=(\ell,m,n)$ is a unit direction vector, $n=\sqrt{1-\ell^2-m^2}$, and $T$ is the sky brightness. The solid-angle measure is $d^2\hat s=d\ell\,dm/n$.
Using the planar Fourier convention $\mathcal F[f](\vec u)\equiv\iint f(\ell,m)e^{-2\pi i(u\ell+vm)}\,d\ell\,dm$, with functions extended by zero outside the visible disk, the integral factorizes as
\begin{equation}
  V_{ij}(\vec u) = \bigl(\tilde T*K_{ij}\bigr)(\vec u),
  \label{eq:visibility_sky_kernel_convolution}
\end{equation}
where $\tilde T=\mathcal F[T/n]$ and $K_{ij}=\mathcal F[\vec E_i\cdot\vec E_j^*]$. This two-dimensional factorization, with a beam kernel independent of $w=b_z/\lambda$, is exact for the coplanar geometry assumed here. For nonzero $w$, it requires either retaining the additional direction-dependent phase or restricting the field sufficiently that this phase can be neglected. The detailed derivation is presented in \cref{app:beam_kernel_derivation}.

\begin{figure}[h!]
  \centering
  \includegraphics[width=\textwidth]{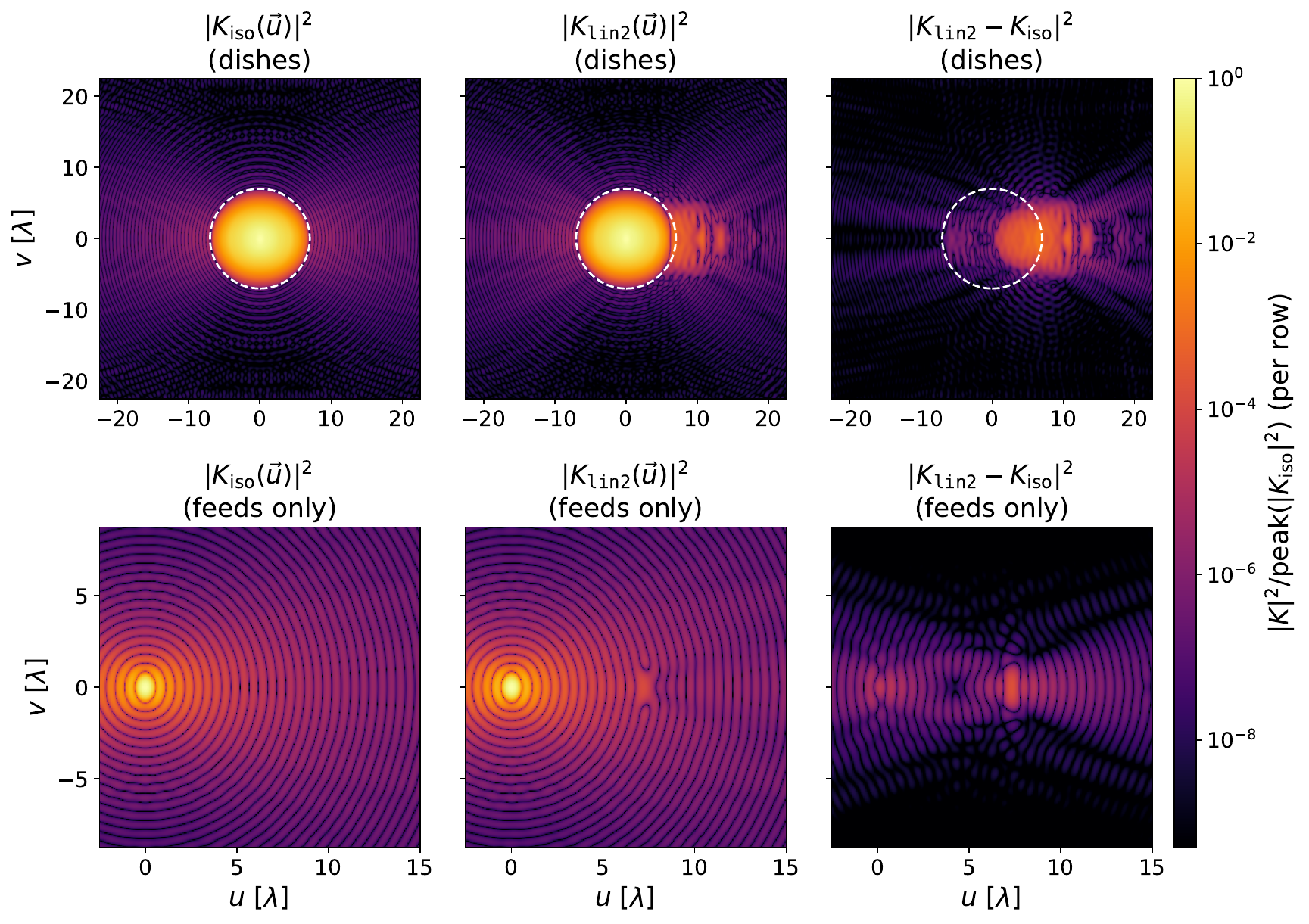}
  \caption{Beam kernels at 150~MHz on the $uv$ plane: the isolated-element kernel power $|K_{\rm iso}(\vec u)|^2$ (left), the coupled-pair kernel power (center), and the power of their complex difference (right), for \texttt{lin2} in the dish model (top row) and the feed-only model (bottom row), each row normalized to its own isolated kernel's peak power. White dashed circles mark the nominal support, of diameter $2D/\lambda$, of the isolated dish autocorrelation.\label{fig:isolated_and_coupled_beam_kernels}}
\end{figure}

The beam kernel describes how the beam product convolves the sky's Fourier modes. At a given frequency, it is a sky-independent complex pattern on the $uv$ plane. For identical isolated dishes, it peaks at $\vec u=0$ and falls off over a scale set by $D/\lambda$, where $D$ is the dish diameter. The beam kernel has been defined previously, for instance, as the \emph{beam power pattern in $uv$ space} in \citet{morales_matejek_2009}, and in fast holographic deconvolution literature as the $uv$-beam \citep{sullivan_et_al_2012,barry_et_al_2019}. This factorization places the sky in $\tilde T$ and the antenna responses, including coupling, in $K_{ij}$. In aperture-plane terms, the kernel is proportional to the cross-correlation of the two elements' aperture fields, summed over the two E-field components. Writing each field as the isolated response plus a perturbation gives
\begin{equation}
\begin{aligned}
  K_{ij}(\vec u)&= \frac{1}{\lambda^2}\int \vec A_i(\vec x)\cdot\vec A_j^*(\vec x-\lambda\vec u)\,d^2x\\
  &= K_{\rm iso}(\vec u)+ \frac{1}{\lambda^2}\int\Big[\vec A_{\rm iso}(\vec x)\cdot\Delta\vec A_j^*(\vec x-\lambda\vec u)\\
  &\qquad+\Delta\vec A_i(\vec x)\cdot\vec A_{\rm iso}^*(\vec x-\lambda\vec u)+\Delta\vec A_i(\vec x)\cdot\Delta\vec A_j^*(\vec x-\lambda\vec u)\Big]\,d^2x.
\end{aligned}
\label{eq:aperture_cross_correlation_kernel}
\end{equation}
where we omit the mode decomposition for simplicity (restored in \cref{subsec:kernel_mode_decomposition}). \Cref{eq:aperture_cross_correlation_kernel} reduces to the autocorrelation of $\vec{A}_{\rm iso}$ for the simple case of two identical, uncoupled antennas. Mode subscripts on kernels correspond to those on aperture fields throughout: $K_{\rm iso}$ for the isolated autocorrelation, $K_{\rm tot}$ for a fully coupled pair, $K_{\rm am}$ and $K_{\rm sm}$ for synthetic beams carrying the corresponding coupling mode; an unsubscripted $K_{ij}$ refers to generic elements $i$ and $j$. \Cref{fig:isolated_and_coupled_beam_kernels} shows kernels for \texttt{lin2}, in both the dish and feed-only models: the isolated kernel $K_{\rm iso}$, the coupled pair's $K_{\texttt{lin2}}$, and their difference (mode decomposition omitted). The feature changes caused by coupling alter which Fourier modes of the sky each baseline is sensitive to compared to the isolated case. We avoid proposing an intuition for the predicted effect of those features on visibilities, as the affected modes depend on each feature's position relative to the sampled baseline, and their contribution to the visibility depends on the observed sky. We do note that coupling tends to spread the beam kernel's power over a wider region of $uv$ space, which can introduce visibility power at fringe rates not expected from the isolated response and produce image artifacts on finer angular scales.

\subsection[Mode decomposition and correction in uv space]{Mode decomposition and correction in $uv$ space}
\label{subsec:kernel_mode_decomposition}

\begin{figure}[h!]
  \centering
  \includegraphics[width=\textwidth]{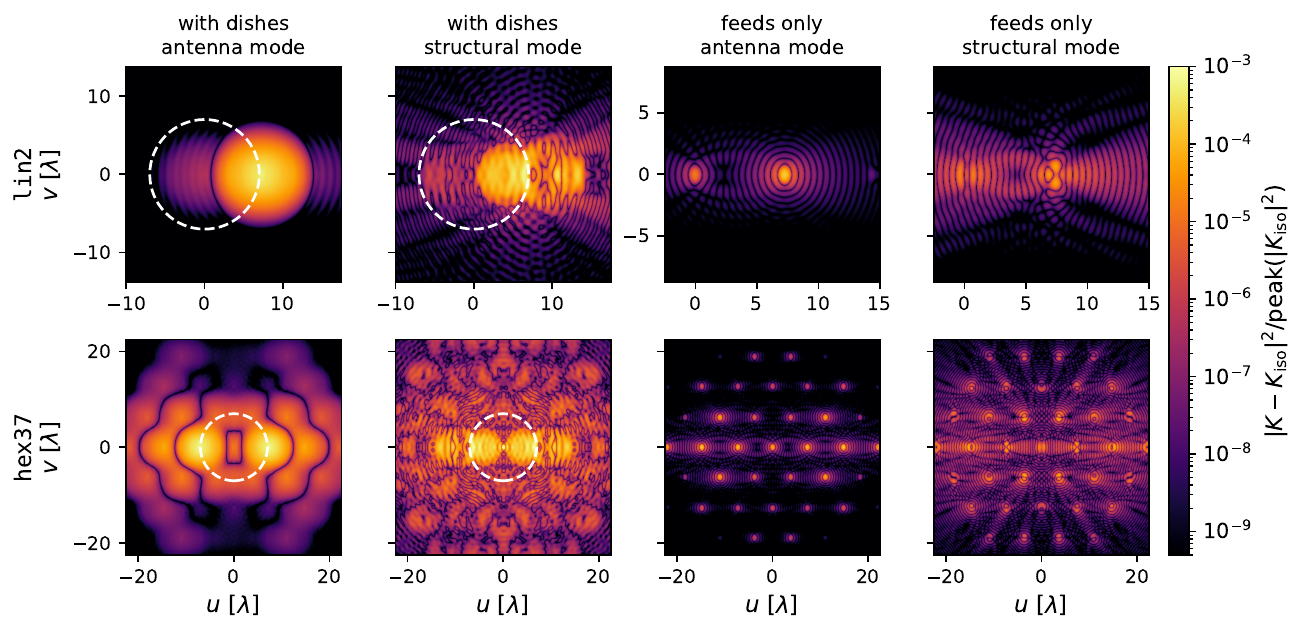}
  \caption{Beam kernel view of the two coupling modes at 150~MHz. Rows: \texttt{lin2} (top) and \texttt{hex37} (bottom). Columns: antenna and structural modes with dishes (left two), then the corresponding feed-only modes (right two). Each panel shows the kernel of a synthetic beam carrying the isolated response plus only the indicated mode, with the complex isolated kernel subtracted; the power is normalized to the corresponding isolated kernel's peak power. The \texttt{lin2} panels use each element's own embedded beam, whereas the \texttt{hex37} panels autocorrelate the interior (central) element's embedded beam, because per-element simulations are too computationally costly. White dashed circles mark the nominal support, of diameter $2D$, of the isolated dish autocorrelation in the dish columns.}\label{fig:antenna_and_structural_beam_kernels}
\end{figure}

To carry the decomposition of \cref{sec:antenna_structural_decomposition} into kernel space, we write
\begin{equation}
\begin{aligned}
  \vec A_{i}(\vec x) &= \vec A_{\rm iso}(\vec x) \;+\; \Delta\vec A_{{\rm am},i}(\vec x) \;+\; \Delta\vec A_{{\rm sm},i}(\vec x)\\
  &=\sum_{k=1}^{N} M_{ik}\, \vec A_{\rm iso}(\vec x - \vec b_{ik}) \;+\; \Delta\vec A_{{\rm sm},i}(\vec x)
\end{aligned}
\label{eq:array_aperture_mixing_and_structural_residual}
\end{equation}
where, in the second line, we have used the fact that the antenna mode consists of shifted copies of the isolated pattern to group them into a matrix-generated contribution: $M_{ik} = \delta_{ik} + \epsilon_{ik}$, $\vec b_{ik} = \vec x_k - \vec x_i$, the $\epsilon_{ik}$ are the projected coefficients of \cref{eq:array_antenna_mode_template_expansion}, and $\delta_{ik}$ is the Kronecker delta. Substituting \cref{eq:array_aperture_mixing_and_structural_residual} into \cref{eq:aperture_cross_correlation_kernel} produces
\begin{equation}
  K_{ij,\rm tot}(\vec u) = \sum_{k=1}^{N}\sum_{l=1}^{N}M_{ik}M_{jl}^*\,K_{\rm iso}\!\left(\vec u-\frac{\vec b_{ik}-\vec b_{jl}}{\lambda}\right)+C_{ij}^{\rm sm}(\vec u).\label{eq:coupled_kernel_template_expansion}
\end{equation}
The double sum is a lattice of translated copies of the isolated kernel, shifted by $(\vec b_{ik}-\vec b_{jl})/\lambda$ in the $uv$ plane and weighted by products of coupling coefficients. We note that $M_{ii}\approx 1$, so the term with $k=i$ and $l=j$ dominates, producing the bright central feature in \cref{fig:isolated_and_coupled_beam_kernels}. All terms containing the structural-mode field are gathered in
\begin{equation}
\begin{aligned}
  C_{ij}^{\rm sm}(\vec u) = \frac{1}{\lambda^2} \int\Bigg\{&\left[\sum_k M_{ik}\,\vec A_{\rm iso}(\vec x-\vec b_{ik})\right]\cdot\Delta\vec A_{{\rm sm},j}^*(\vec x-\lambda\vec u)\\
  &+\Delta\vec A_{{\rm sm},i}(\vec x)\cdot\left[\sum_l M_{jl}\, \vec A_{\rm iso}(\vec x-\vec b_{jl}-\lambda\vec u)\right]^*\\
  &+\Delta\vec A_{{\rm sm},i}(\vec x)\cdot\Delta\vec A_{{\rm sm},j}^*(\vec x-\lambda\vec u)\Bigg\}\,d^2x.
\end{aligned}
\label{eq:structural_kernel_terms}
\end{equation}
To write the result in matrix form, we include each antenna pair's geometric displacement in its kernel. With $\vec u_{ij}=\vec b_{ij}/\lambda=(\vec x_j-\vec x_i)/\lambda$, we define $[\mathbf K(\vec u)]_{ij}=K_{ij}(\vec u+\vec u_{ij})$ and $[\mathbf C^{\rm sm}(\vec u)]_{ij} =C_{ij}^{\rm sm}(\vec u+\vec u_{ij})$. In particular,
$[\mathbf K_{\rm iso}(\vec u)]_{ij} =K_{\rm iso}(\vec u+\vec u_{ij})$. With this convention, \cref{eq:coupled_kernel_template_expansion} reads
\begin{equation}
  \mathbf K_{\rm tot}(\vec u)
  =\mathbf M\,\mathbf K_{\rm iso}(\vec u)\,
  \mathbf M^\dagger+\mathbf C^{\rm sm}(\vec u).
  \label{eq:kernel_matrix_mixing_and_structural_residual}
\end{equation}
Convolving each matrix entry with $\tilde T$ and evaluating at
$\vec u=0$ gives the corresponding measured visibility, so the visibility matrices obey the same mixing relation, recovering \cref{eq:visibility_mixing_matrix} when $\mathbf C^{\rm sm}=0$.
We suppress the common argument $\vec u$ in the matrix equations below.
\Cref{fig:antenna_and_structural_beam_kernels} illustrates each mode through the kernel of a synthetic beam carrying that mode together with the isolated response, with the isolated kernel subtracted. These two maps need not add to the total perturbation because the full beam product also contains cross terms between the modes.
Applying the inversion then yields
\begin{equation}
  \mathbf M^{-1}\, \mathbf K_{\rm tot}\, \mathbf M^{-\dagger} = \mathbf K_{\rm iso} \;+\; \mathbf M^{-1}\, \mathbf C^{\rm sm}\, \mathbf M^{-\dagger} \;\equiv\; \mathbf K_{\rm iso} + \mathbf K_{{\rm res},\epsilon},
  \label{eq:kernel_residual_after_matrix_inversion}
\end{equation}
The inversion removes the antenna-mode factor exactly and transforms the structural term by $\mathbf M^{-1}$. For purely antenna-mode coupling, $\mathbf C^{\rm sm}$ and $\mathbf K_{{\rm res},\epsilon}$ vanish and the inversion removes all coupling.

The projected coefficients $\epsilon$ minimize the aperture-space residual $r_\epsilon$ (\autoref{eq:aperture_projection_residual_fraction}). There, expressing the coupled response in terms of isolated responses helps define and build an intuition for the antenna and structural modes. We now seek a correction that acts on the coupled responses as it would on measured data. We therefore use the isolated kernel as the fitting target and form the corrected kernel from weighted copies of the coupled kernel,
\begin{equation}
  K_{{\rm corr},\sigma}(\vec u)
  =K_{\rm tot}(\vec u)
  -\sum_{\vec q\in\mathcal Q_{ij}}\sigma_{\vec q}\,
  K_{\rm tot}\!\left(\vec u-\frac{\vec q}{\lambda}\right),
  \label{eq:linear_and_quadratic_visibility_mixing}
\end{equation}
where $\mathcal Q_{ij}=\{\vec b_{ik}\}\cup\{-\vec b_{jl}\}\cup\{\vec b_{ik}-\vec b_{jl}\}$ is the set of physical shifts, in meters, with $k,l$ running over the antennas for each output pair $(i,j)$. This set includes zero separation; the corresponding shifts in the $uv$ plane are $\vec q/\lambda$. The coefficients now fit the correction directly rather than first fitting and then inverting a forward model---fitting a correction operator that plays the role of $\mathbf{\bar{M}}^{-1}$ introduced in \cref{eq:baseline_visibility_mixing_matrix}. We fit the coefficients jointly by minimizing $R_\sigma$, defined below, without requiring them to satisfy the antenna-level product relations inherited from $\mathbf M$. The residual is $\mathbf K_{{\rm res},\sigma}=\mathbf K_{{\rm corr},\sigma}-\mathbf K_{\rm iso}$, where $\mathbf K_{{\rm corr},\sigma}$ collects the corrected pair kernels using the same geometric-shift convention as above. We compare the two corrections through
\begin{equation}
 R_\epsilon(\nu)\equiv
 \frac{\lVert\mathbf K_{{\rm res},\epsilon}\rVert^2}
 {\lVert\mathbf K_{\rm tot}-\mathbf K_{\rm iso}\rVert^2},
 \qquad
 R_\sigma(\nu)\equiv
 \frac{\lVert\mathbf K_{{\rm res},\sigma}\rVert^2}
 {\lVert\mathbf K_{\rm tot}-\mathbf K_{\rm iso}\rVert^2},
 \label{eq:kernel_correction_residual_fractions}
\end{equation}
using the same norm over the $uv$ plane and antenna pairs.
At each frequency, the $\sigma$ fit minimizes the numerator of $R_\sigma$. In contrast, $R_\epsilon$ evaluates the correction obtained from the aperture-space fit. These ratios measure the remaining beam-kernel perturbation power: zero denotes complete removal, unity no net suppression, and values above unity amplification. Here beam-kernel power means the integrated squared magnitude,
\[
  \lVert\mathbf K\rVert^2
  \equiv \sum_{i,j}\iint
  \left|[\mathbf K(\vec u)]_{ij}\right|^2\,du\,dv.
\]
The visibility and power spectrum residuals also depend on sky convolution and baseline sampling, and we revisit the kernel fit accordingly in \cref{sec:coupling_in_simulated_visibilities}.

\begin{figure}[h!]
  \centering
  \includegraphics[width=\textwidth]{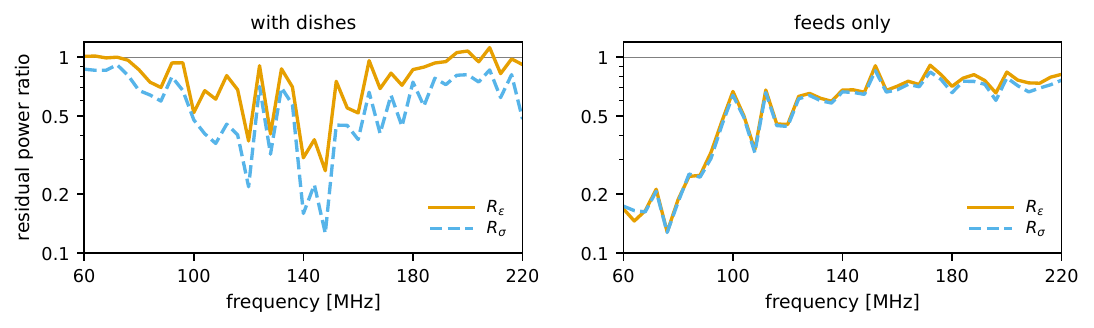}
  \caption{Residual power ratios in $uv$ space for the \texttt{hex37} array in the identical-element idealization, for the dish (left) and feed-only (right) models. Solid curves show $R_\epsilon$ after coupling-matrix inversion with the projected coefficients $\epsilon$ fitted in aperture space; dashed curves show $R_\sigma$ after the free-weight projection, fitted in $uv$ space (both defined in \cref{eq:kernel_correction_residual_fractions}). For dishes, the minimum near $140$--$150$~MHz coincides with a relatively strong perturbation that the corrections capture well; it is a minimum in the fractional residual, not necessarily in the absolute remaining power. The rising feed-only residual follows the increasing structural-mode fraction in \cref{fig:aperture_mode_power_budget}, consistent with the feed becoming electrically larger. The gray line at unity marks no net suppression.}\label{fig:matrix_inversion_and_free_weight_kernel_residuals}
\end{figure}

\Cref{fig:matrix_inversion_and_free_weight_kernel_residuals} compares the two corrections for the \texttt{hex37} dish and feed-only models. Although the residual of the $\epsilon$ correction originates in the structural-mode term, forming beam products does not preserve the aperture-space orthogonality, and thus the kernel-space ratio $R_\epsilon$ need not equal the aperture-space residual $r_\epsilon$. The inversion generally leaves a kernel-space residual fraction below $r_\epsilon$, although it amplifies the perturbation at some frequencies.

This favorable behavior can be understood from the operation that produces the perturbed kernel: to first order, $\Delta\vec{A}$ is cross-correlated with $\vec{A}_{\rm iso}$ (\cref{eq:aperture_cross_correlation_kernel}), so its contribution depends on how strongly it correlates with the isolated response. The antenna mode, consisting of translated copies of $\vec{A}_{\rm iso}$, produces translated copies of $K_{\rm iso}$. In contrast, the structural mode's asymmetric sub-dish features can correlate less strongly with the isolated response. This is emphasized in the dish panels of \Cref{fig:antenna_and_structural_beam_kernels}, where the antenna-mode perturbation in kernel space has a brightness similar to that of the structural mode, even though its aperture power budget is roughly two orders of magnitude smaller.

\subsection{Enrichment with fractional baselines}
\label{subsec:kernel_template_enrichment}

\begin{figure}[h!]
  \centering
  \includegraphics[width=\textwidth]{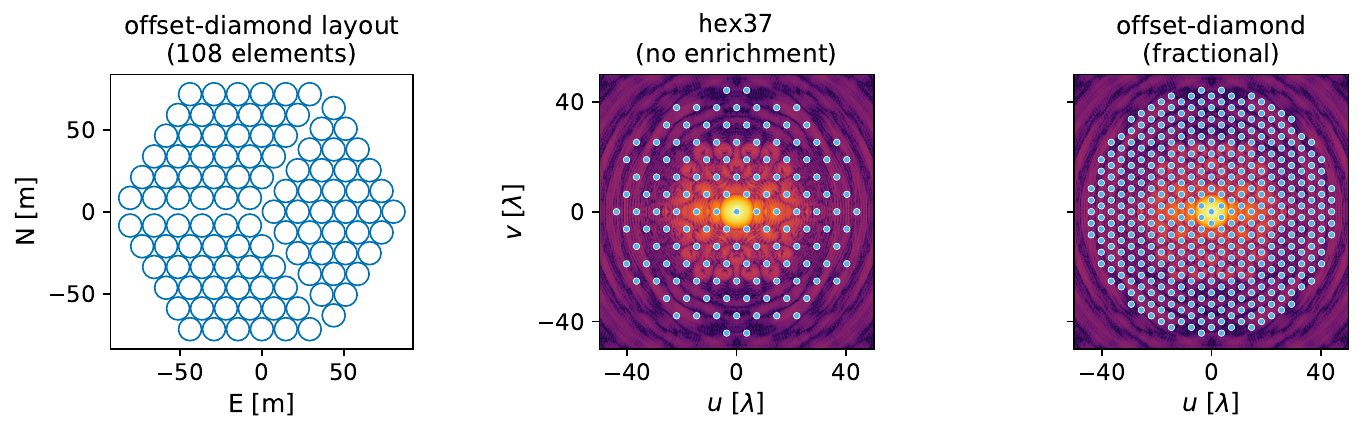}
  \caption{Left: the 108-element offset-diamond layout used to generate fractional shifts. Center and right: template centers $\vec q/\lambda$ derived from the \texttt{hex37} and offset-diamond layouts, overlaid on the fully coupled \texttt{hex37} dish beam-kernel power at 150~MHz. The antenna layout uses physical coordinates in meters; the beam-kernel panels use $uv$ coordinates. The dots locate correction templates, rather than measurements of the sky-convolved kernel.}\label{fig:enriched_kernel_template_geometries}
\end{figure}

The template set of \cref{eq:linear_and_quadratic_visibility_mixing} uses shifts derived from the \texttt{hex37} layout. We test whether adding intermediate shifts improves the correction by using an offset-diamond layout inspired by HERA's split core \citep{dillon_parsons_2016}. Offsetting sectors by fractions of the lattice vectors introduces \emph{fractional baselines}, and hence template centers between those of the original hexagonal lattice. We construct the enlarged shift set from this layout while retaining the same coupled \texttt{hex37} beam and minimizing the same full-plane residual $R_\sigma$. This comparison therefore tests the effect of the available template positions, rather than the change in the embedded beam that would accompany a different physical array. \Cref{fig:enriched_kernel_template_geometries} shows the layout and the two shift sets.

\begin{figure}[h!]
  \centering
  \includegraphics[width=\textwidth]{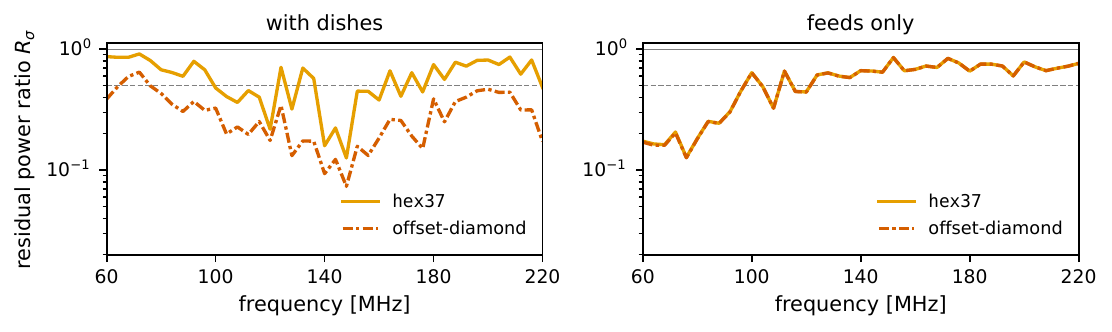}
  \caption{Beam-kernel residual power ratios $R_\sigma$ (\cref{eq:kernel_correction_residual_fractions}) for template shifts derived from the \texttt{hex37} and offset-diamond layouts shown in \cref{fig:enriched_kernel_template_geometries}, for dishes (left) and feeds only (right). Both fits use the same coupled \texttt{hex37} beam and minimize the residual over the full $uv$ plane. The gray lines mark no net suppression (solid, unity) and half of the coupling power remaining (dashed, $0.5$). Fractional shifts substantially improve suppression for dishes but provide little improvement for feeds, showing that their usefulness depends on the spatial structure of the beam-kernel perturbation.}\label{fig:enriched_kernel_projection_residuals}
\end{figure}

\Cref{fig:enriched_kernel_projection_residuals} compares the two template sets. For dishes, the fractional shifts reduce the beam-kernel residual, consistent with intermediate template positions helping to represent the extended structure associated with the dish rims. For feeds, the two curves nearly coincide: the additional positions provide little improvement for the more compact perturbation features, likely because the feed's structural mode does not lie at the location provided by the additional baselines. The benefit of fractional baselines therefore depends on the element response with respect to the location of those baselines, rather than on the size of the template set alone.

\section{Coupling modes and correction in simulated data}
\label{sec:coupling_in_simulated_visibilities}

In \cref{sec:coupling_in_kernel_space}, we saw that forming beam products changes the relative importance of the coupling modes, so the aperture-space power budgets do not directly predict the kernel-space residuals. We expect a similar change when going from kernels to visibilities: the kernel fits minimize a norm uniformly over the plane, whereas visibilities sample the sky-convolved kernel at the array's baseline locations. We now follow this step in simulated observations, examining the mode imprints and comparing corrections derived from the beams (aperture or kernel space) with fits made directly to the visibilities.

\subsection[Coupling modes in visibility space]{Coupling modes in visibility space}
\label{subsec:visibility_and_image_mode_simulations}

We run noiseless visibility simulations with isolated, fully coupled, and synthetically mode-separated beams, taking the fully coupled beam from the central element of the \texttt{hex37} electromagnetic simulation. Unless otherwise specified, the simulated antenna positions also follow the \texttt{hex37} layout ($63$ unique nonzero baselines), with the same beam assigned to every element, including those at the edge, as the computational costs of using distinct primary beams were prohibitive. This identical-element approximation makes the array perfectly redundant, neglecting the differences in coupling between antennas with different neighborhoods in a real array (see \cref{subsec:requirements_for_practical_correction}), and parallels the autocorrelation approach used to form the \texttt{hex37} kernels in \Cref{fig:antenna_and_structural_beam_kernels}. We construct the synthetic mode-separated beams using the same projected decomposition as in \cref{subsec:array_aperture_mode_decomposition}.

The simulated sky combines the GLEAM catalog \citep{hurley_walker_et_al_2017} with the GSM (\texttt{nside} = 256) diffuse model \citep{de_oliveira_costa_et_al_2008}. We interpolate the complex beams to $147$ channels between $141$ and $159$~MHz and use \texttt{fftvis} \citep{cox_et_al_2025} to simulate observations. The observations use HERA's site (latitude $-30.7^\circ$, longitude $21.4^\circ$) and cover LST $4^{\rm h}$--$6^{\rm h}$ in $120$ integrations at one-minute cadence. We refer to this as the fiducial field; it is characteristic of the kinds of quiet fields---without either a bright point source or the Galactic plane near zenith---that telescopes such as HERA tend to target.

\begin{figure}[h!]
  \centering
  \includegraphics[width=\textwidth]{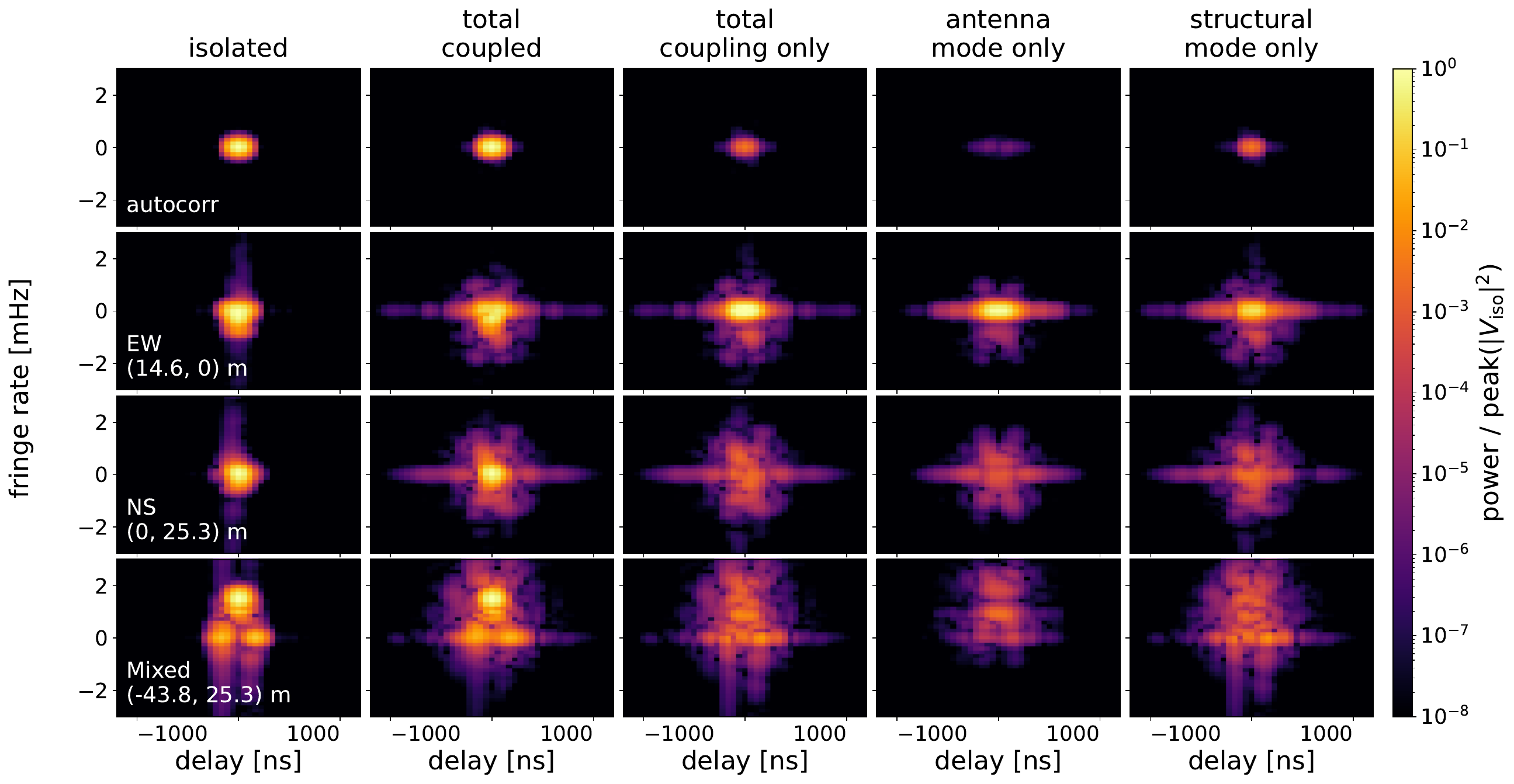}
  \caption{The fringe-rate/delay representation of simulated visibilities, for the autocorrelation (top row) and three representative baselines (lower rows), each row normalized to its isolated case's peak power. Columns: isolated, fully coupled, total coupling only, antenna mode only, structural mode only (the last three with the isolated visibilities subtracted).\label{fig:visibility_modes_in_fringe_rate_delay_space}}
\end{figure}

\Cref{fig:visibility_modes_in_fringe_rate_delay_space} shows the autocorrelation and three baselines in fringe-rate and delay space for the fiducial field. The X-like feature described by \citet{rath_et_al_2025} is visible, albeit faintly, especially in the antenna-mode column; the structural contribution is more smeared out. This feature would be more visible in a larger simulated array, as longer baselines extend the X.

\subsection[Beam-derived and visibility-fitted corrections]{Beam-derived and visibility-fitted corrections}
\label{subsec:beam_and_visibility_fitted_corrections}
Following the kernel-space analysis of \Cref{sec:coupling_in_kernel_space}, we now quantify how much coupling the same corrections remove after sky convolution and baseline sampling. We first consider beam-derived, sky-agnostic corrections, whose coefficients are determined solely from the isolated and coupled beams. In other words, we take the coefficients found in \Cref{sec:coupling_in_kernel_space} and apply them to visibilities instead of kernels.

We also test a sky-agnostic correction that prioritizes the 63 unique nonzero baselines sampled by \texttt{hex37}. For this fit, we retain solid-angle weighting when transforming the beam product, defining
\[
  K^\Omega_{ij}(\vec u)
  \equiv \mathcal F\!\left[
  \frac{\vec E_i\cdot\vec E_j^*}{n}\right](\vec u).
\]
This differs from the planar beam kernel of \cref{subsec:beam_kernel_definition} and is equivalent to the visibility response to a sky of uniform brightness. At each frequency, we minimize
\[
  \sum_{\vec u\in\mathcal B_\nu}
  \left|K^\Omega_{{\rm corr},\sigma}(\vec u)
  -K^\Omega_{\rm iso}(\vec u)\right|^2,
\]
where $\mathcal B_\nu$ contains the sampled positions $\vec b/\lambda$, and the corrected response uses the shifted-template form of \cref{eq:linear_and_quadratic_visibility_mixing}. The fit uses no information about the observed sky. We label it ``$\sigma$ (baseline points)'' and the full-plane fit ``$\sigma$ (uniform)''.

This is an underdetermined interpolation, with 139 complex coefficients constrained by 63 complex samples; we select the solution with the smallest coefficient norm. A small residual at those points does not constrain the response elsewhere, including near the origin where the isolated response is brightest. It therefore need not imply an accurate correction for an arbitrary sky. We assess the resulting coefficients by applying them to the simulated visibilities, where sky weighting determines which parts of the beam response contribute.

\begin{figure}[h!]
  \centering
  \includegraphics[width=\textwidth]{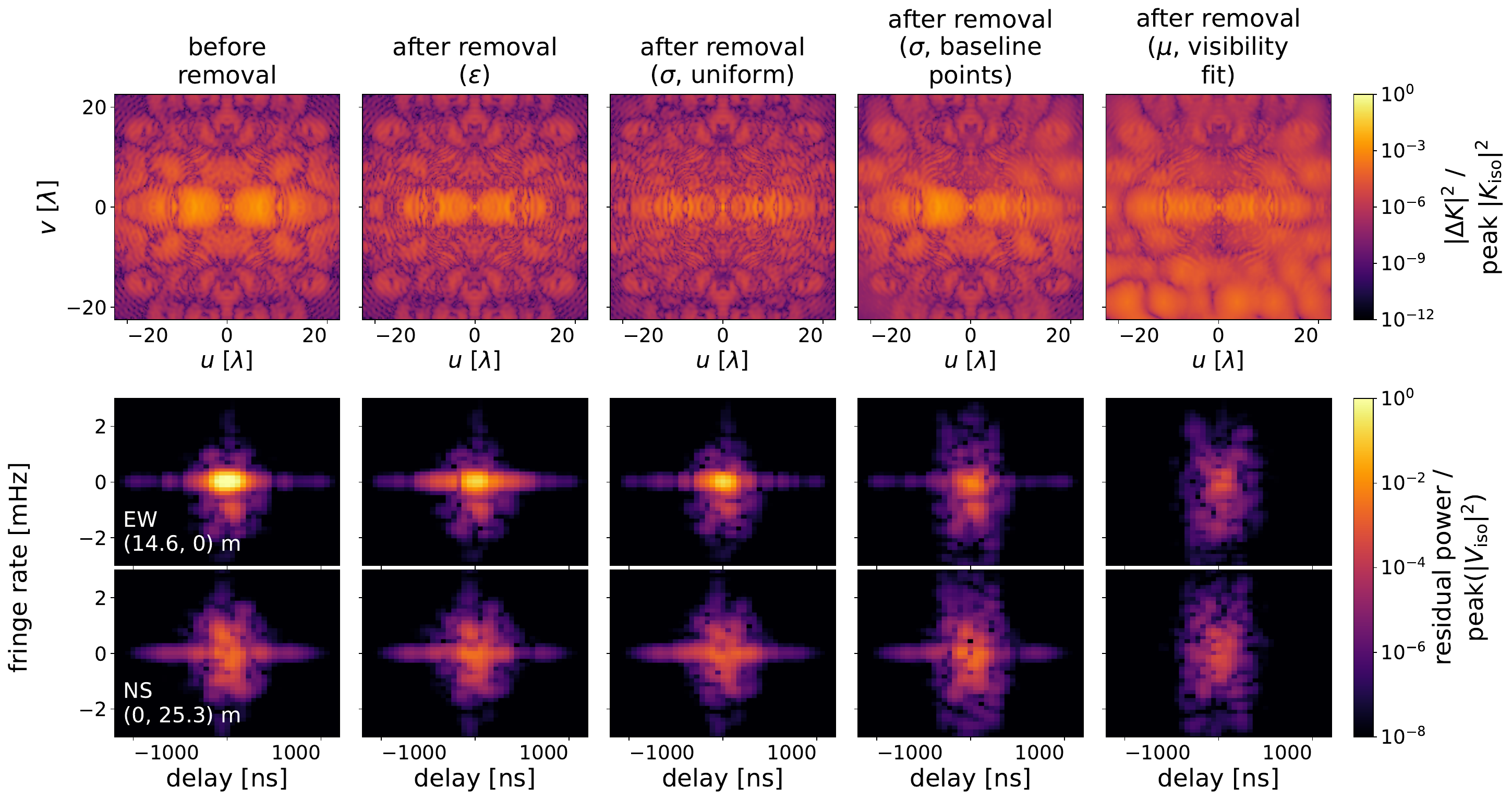}
  \caption{Coupling corrections in $uv$ and visibility space for the simulations of \cref{subsec:visibility_and_image_mode_simulations}. Top row: squared beam-kernel residual at $150$~MHz, normalized to the isolated kernel's peak power; all panels use the planar beam-kernel definition of \cref{subsec:beam_kernel_definition}. Middle and bottom rows: fringe-rate/delay power of corrected minus isolated visibilities at the shortest EW and shortest pure NS baselines, each row normalized to the peak fringe-rate/delay power of its own isolated visibility. Columns: no correction, projected $\epsilon$, $\sigma$ fitted uniformly over the kernel's $uv$ plane, $\sigma$ fitted at the baseline points, and $\mu$ fitted directly to known isolated and coupled visibilities (\cref{eq:visibility_template_fit}). The $\epsilon$ column applies the inverse of the projected coupling model; the $\sigma$ and $\mu$ columns use independently fitted coefficients on coupled templates. The final column has small visibility residuals but a large beam-kernel residual. The $\mu$ fit uses only measured visibilities, minimizing the residual after sky convolution and baseline sampling. Applying those coefficients to the full beam kernel therefore need not reduce its residual throughout the $uv$ plane.}\label{fig:kernel_and_visibility_correction_comparison}
\end{figure}

\begin{figure}[h!]
  \centering
  \includegraphics[width=\textwidth]{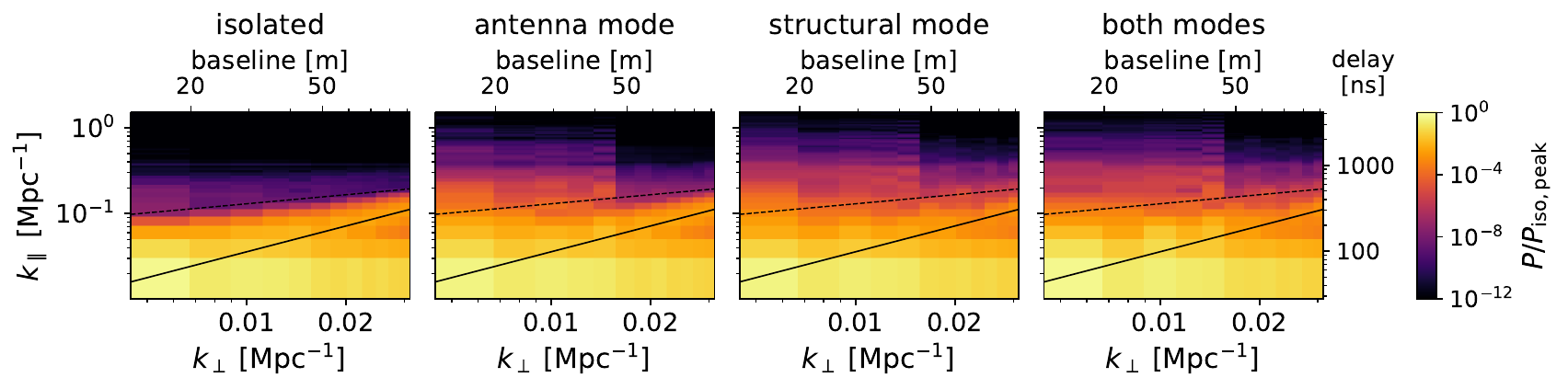}
  \caption{Cylindrically binned power spectra of the simulated visibilities. From left to right: isolated, isolated plus antenna-mode coupling, isolated plus structural-mode coupling, and fully coupled (both modes). The isolated contribution is retained in all four panels, and all spectra are normalized to the isolated peak power. The solid line marks the horizon delay; the dashed line marks the horizon plus the taper main-lobe half-width $4/B\approx220$~ns. Coupling adds power beyond the horizon, with a marked decline near a baseline length of 46~m in the antenna-mode and fully coupled cases, consistent with the finite spatial extent of the electromagnetic model used to construct the coupled beam.}\label{fig:uncorrected_visibility_power_spectra}
\end{figure}

\begin{figure}[p]
  \centering
  \includegraphics[width=\textwidth]{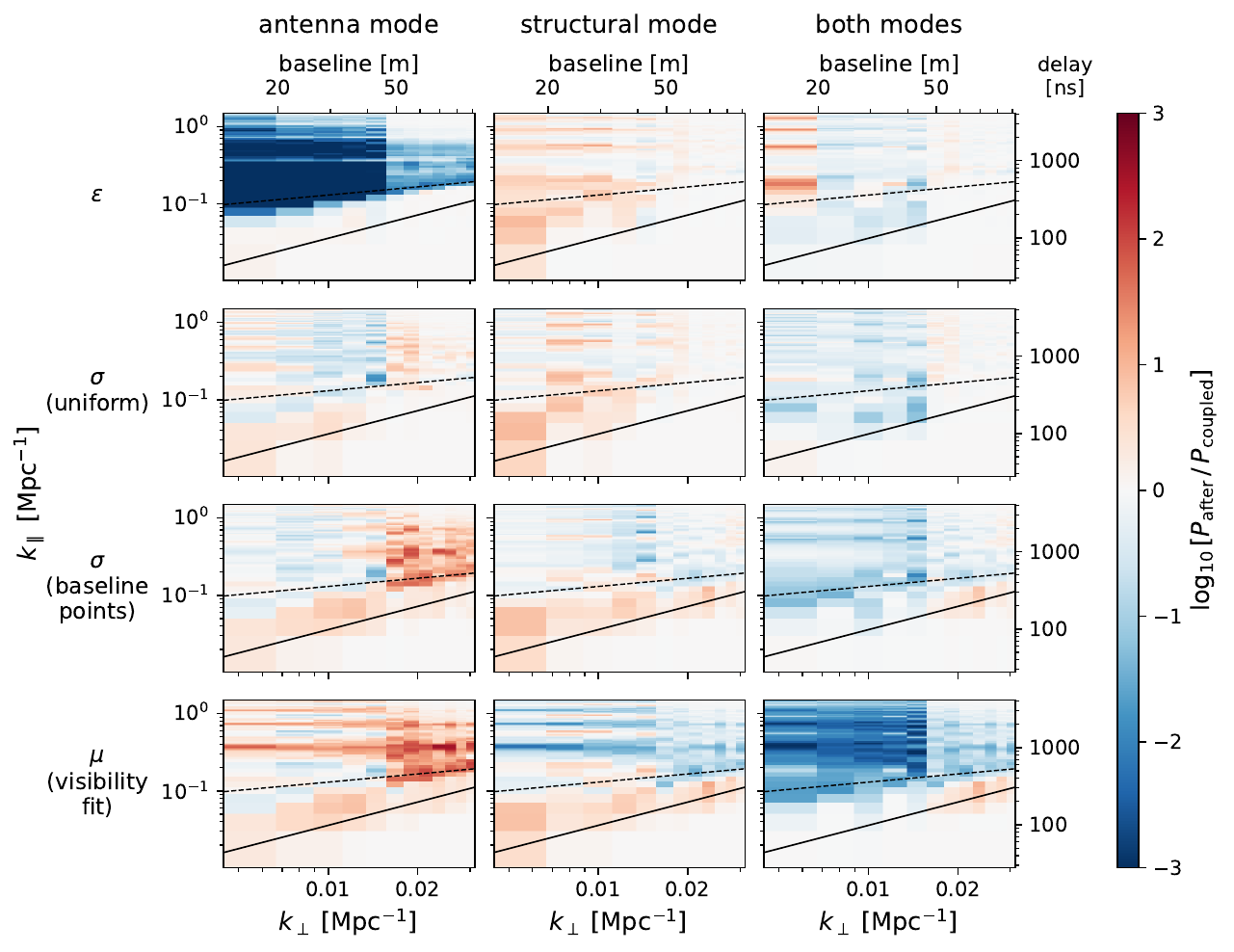}
  \caption{Effects of the corrections on the power spectra of \cref{fig:uncorrected_visibility_power_spectra}. Columns: visibilities contaminated by the antenna mode alone, the structural mode alone, and both modes. Rows: projected $\epsilon$, uniform-kernel $\sigma$, baseline-point $\sigma$, and $\mu$ fitted to known isolated visibilities. Each correction is determined from the fully coupled case and applied unchanged to all three cases. Colors show $\log_{10}(P_{\rm after}/P_{\rm coupled})$: white means no change, while saturated blue and red denote decreases and increases of at least a factor of $1000$, respectively. The horizon and taper-buffer lines are as in \cref{fig:uncorrected_visibility_power_spectra}. The broad blue regions show strong suppression, notably in the antenna-mode-only case for the $\epsilon$ inversion and in the fully coupled case for the $\mu$ fit. Red regions indicate increased power, as when the $\mu$ correction is applied to the antenna-mode-only visibilities: a correction optimized for both modes together need not suppress each mode separately.}\label{fig:corrected_visibility_power_spectra}
\end{figure}

We compare these beam-derived corrections with a fit performed directly in visibility space, using the simulated isolated visibilities as a known target. This measures the suppression attainable for the specified sky and template set. In observations, that target is generally unknown, and estimating it jointly with the correction introduces the sky-coupling degeneracy discussed in \cref{sec:discussion}. As with the $\sigma$ fit of \cref{eq:linear_and_quadratic_visibility_mixing}, we directly fit a baseline-level correction of the $\mathbf{\bar{M}}^{-1}$ form introduced with \cref{eq:baseline_visibility_mixing_matrix},
\begin{equation}\label{eq:visibility_template_fit}
  V_{\rm corr}(\vec u,t,\nu)
  =V_{\rm cpl}(\vec u,t,\nu)
  -\sum_{\vec q}\mu_{\vec q}(\nu)\,
  V_{\rm cpl}\!\left(
  \vec u-\frac{\vec q}{\lambda},t,\nu
  \right),
\end{equation}
where $\lambda=c/\nu$ and the same coefficient set is used across all integrations and output baselines, including only terms for which the input visibility at $\vec u-\vec q/\lambda$ is available. The $\mu$ coefficients act on the visibilities just as the $\sigma$ coefficients do, but we use a different symbol to identify the space in which the fit is performed. All $\sigma$ and $\mu$ fits in this subsection use the \texttt{hex37} template set, without fractional-baseline enrichment. \Cref{fig:uncorrected_visibility_power_spectra} shows the uncorrected power spectra; the corrections' effects on visibilities and power spectra are shown in \cref{fig:kernel_and_visibility_correction_comparison,fig:corrected_visibility_power_spectra}.

\Cref{fig:kernel_and_visibility_correction_comparison} shows that suppression in $uv$ space does not translate directly into suppression of the visibilities. The $\mu$ fit gives the faintest visibility residuals despite leaving a relatively bright full-plane beam-kernel residual, while the baseline-point $\sigma$ correction gives an intermediate result. The power spectra in \cref{fig:corrected_visibility_power_spectra} show a related dependence on the fitted target: the baseline-point $\sigma$ and $\mu$ corrections suppress the fully coupled case strongly but can increase power when applied to the antenna-mode-only case. The projected $\epsilon$ inverse is particularly effective for that mode, as expected from the antenna-level model; the uniform $\sigma$ fit can also correct part of the structural-mode contribution.

The coupled spectra in \cref{fig:uncorrected_visibility_power_spectra} show a marked decline near a baseline length of 46~m, consistent with the finite spatial extent of the electromagnetic model. In \cref{eq:aperture_cross_correlation_kernel}, the terms linear in $\Delta\vec A$ correlate a perturbation with the isolated aperture, whereas the quadratic term correlates perturbations at different neighbors and can extend to larger separations. These are characteristic scales rather than sharp support boundaries: the reconstructed aperture fields contain Fourier-truncation tails, and visibility formation additionally convolves the beam kernel with the sky.

\begin{figure}[h!]
  \centering
  \includegraphics[width=\textwidth]{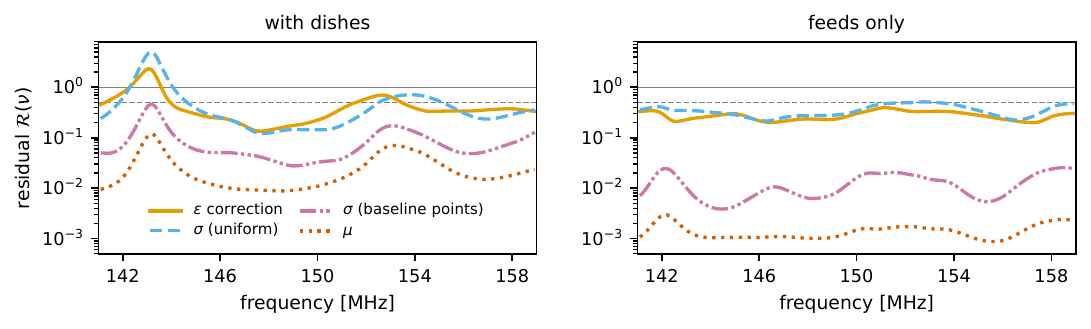}
  \caption{Visibility-correction residual fractions $\mathcal{R}$ (\cref{eq:visibility_space_residual}) with dish (left) and feed-only (right) beams, for the fiducial field. Residual power is summed over integrations and baselines and normalized by the uncorrected coupling power on the same samples at each frequency. All curves use the 63 nonzero \texttt{hex37} baselines. Gray lines mark no net suppression (solid, unity) and half of the coupling power remaining (dashed, $0.5$). The dish-panel corrections are the same as in \cref{fig:kernel_and_visibility_correction_comparison,fig:corrected_visibility_power_spectra}. Fitting at the baseline points improves substantially on the uniform kernel fit, especially for feeds, while fitting directly to visibilities gives the smallest residuals in both models. The dish-panel peak near $143$~MHz coincides with a trough in the uncorrected coupling power used as the denominator.}
  \label{fig:visibility_correction_residuals}
\end{figure}

We now compare all removal methods introduced so far with a visibility-space residual,
\begin{equation}\label{eq:visibility_space_residual}
  \mathcal{R}(\nu) = \frac{\lVert \mathbf{V}_{\rm corr}-\mathbf{V}_{\rm iso}\rVert^2}{\lVert \mathbf{V}_{\rm cpl}-\mathbf{V}_{\rm iso}\rVert ^2},
\end{equation}
where the squared norm sums over integrations and the 63 unique nonzero baselines at each frequency, without weighting by the number of redundant antenna pairs. The $\mu$ coefficients are obtained by minimizing this $\mathcal{R}(\nu)$ independently at each frequency; its denominator is fixed by the simulated coupled and isolated visibilities. This residual power ratio is analogous to $R_{\epsilon,\sigma}$ in \Cref{subsec:kernel_mode_decomposition}. \Cref{fig:visibility_correction_residuals} shows these residuals for the dish and feed-only simulations. The peak above unity near 143~MHz corresponds to a deep trough in the coupling power in visibility space, more sensitive to inadequate removal due to performing the fits in other spaces. The aperture-space correction and uniform kernel fit result in relatively poor subtraction---less than one order of magnitude---while fitting the coefficients directly to the visibilities reaches much deeper in both models. The baseline-point $\sigma$ fit improves substantially on the uniform fit, especially in the feed-only setup, but still leaves larger residuals than the visibility-space fit.

\begin{figure}[h!]
  \centering
  \includegraphics[width=\textwidth]{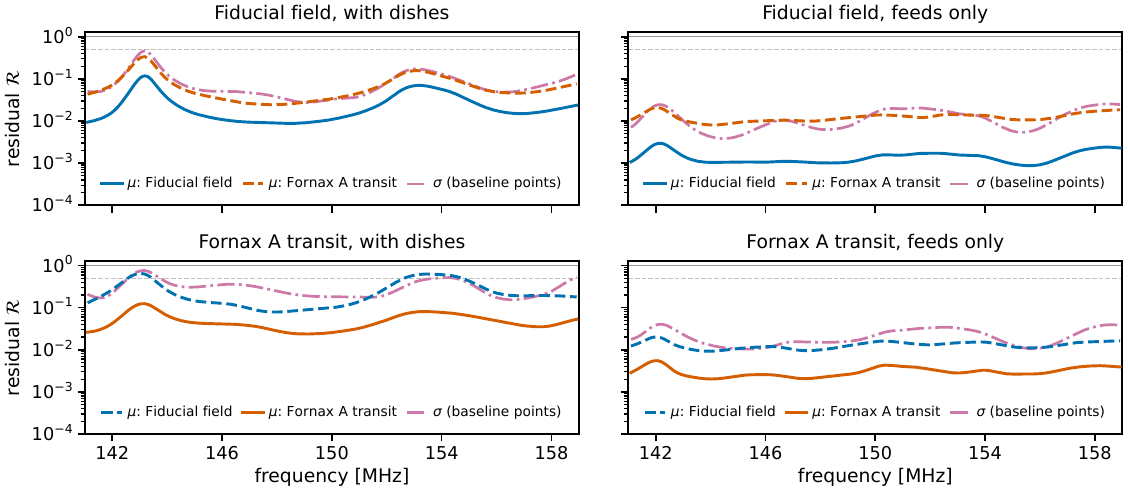}
  \caption{Transfer of $\mu$ coefficients between sky fields, for dish (left) and feed-only (right) beams. The top row repeats the fiducial field $\mu$ and baseline-point $\sigma$ curves from \cref{fig:visibility_correction_residuals}, adding the correction obtained with $\mu$ fitted on Fornax~A; the bottom row performs the reverse test on Fornax~A. Colors identify the field used to fit $\mu$: solid curves use the target field's own fit, while dashed curves transfer the other field's coefficients without refitting. The pink dash-dotted curve shows baseline-point $\sigma$. Each $\mathcal{R}$ uses the 63 \texttt{hex37} baselines and the integrations and normalization of the field being corrected (\cref{eq:visibility_space_residual}); gray lines mark unity and $0.5$. Transferred fits still suppress coupling, but generally leave residuals comparable to the sky-agnostic baseline-point correction. The loss of suppression despite unchanged beam models shows that the optimal visibility-fit coefficients depend on the illuminating sky. The smallest residuals are measured on the same samples used to fit the coefficients and do not demonstrate equally accurate correction of a different sky.}
  \label{fig:visibility_coefficient_field_transfer}
\end{figure}

Since the $\mu$ coefficients are fitted after convolution with a particular sky, we also test whether they transfer between fields. We simulate a field centered on Fornax~A with the same \texttt{hex37} layout, beam models, and frequency channels, using 150 one-minute integrations starting at LST $2.128^{\rm h}$ and latitude $-37.21^\circ$ to bring Fornax~A through zenith; while the fiducial field was mostly diffuse, the Fornax~A transit is dominated by a bright central point source, and thus provides a qualitatively different illumination. In \cref{fig:visibility_coefficient_field_transfer}, we apply each field's $\mu$ coefficients unchanged to the other field's coupled visibilities. The transferred corrections still suppress coupling, but leave larger residuals than a fit to the field being corrected, generally comparable to the sky-agnostic baseline-point $\sigma$ correction. The additional suppression obtained on the fitted field therefore depends on its particular illumination, rather than representing a universally accurate inverse of the instrumental response. This limited generalization is consistent with overfitting to the fitted sky and with a correction model that only approximates coupling: its free coefficients can compensate for model mismatch differently under different illuminations. The fitted correction coefficients can consequently vary with LST even when the underlying instrument is stable.

In visibility space, we restrict our comparisons to the 
\texttt{hex37} template set. Not only does it already give deeper overall suppression than the sky-agnostic corrections tested here---so long as the fit is performed on the field being corrected---but as mentioned previously, deploying this approach to observations raises an underdetermination problem: the isolated sky signal used as the fitting target is generally unknown and must be inferred along with the coupling coefficients. Enlarging the template set with the offset-diamond layout introduces further unknown coefficients, increasing the freedom to absorb sky-model errors or cosmological emission into the correction. Assessing the practical benefit of those added baselines therefore requires a method for constraining those coefficients and tests of signal preservation, which we leave to future work. We discuss possible constraints in \cref{sec:discussion}.

\clearpage 
\section[Discussion]{Discussion}
\label{sec:discussion}

We presented two broad families of corrections: sky-agnostic corrections, which required knowledge of the isolated and coupled beams (aperture and kernel spaces; \cref{sec:antenna_structural_decomposition,sec:coupling_in_kernel_space}), and sky-informed corrections, which additionally required knowledge of the sky (visibility space; \cref{sec:coupling_in_simulated_visibilities}). It is generally difficult but not impossible to obtain beam models to varying degrees of accuracy via simulations or measurements, so the sky-agnostic corrections may be feasible for some arrays. Their suppression is, however, limited; the best-performing sky-agnostic fit, where the residuals are minimized at the precise locations of the sampled baselines, reduces the coupling power by approximately one order of magnitude in the setup that includes dishes, and two in the feed-only case for the fiducial field.

The sky-informed fits give smaller residuals on the field used to determine their coefficients, but their reduced accuracy on another field limits the interpretation of that improvement. They demonstrate that linear combinations of the available coupled visibilities can closely reproduce the known isolated visibilities for a specified sky. They do not establish a sky-independent correction or a practical estimator when the isolated sky is unknown. We outline below several ways to constrain the sky or coupling parameters; these could reduce overfitting and help identify a correction from observations, but restrictions on the correction model may also exclude combinations of coupled responses that contribute to the small residuals obtained with freely fitted $\sigma$ and $\mu$ coefficients, sacrificing some of the suppression.

\subsection[Constraining the sky-coupling degeneracy]{Constraining the sky-coupling degeneracy}
\label{subsec:instrument_and_sky_constraints}

The number of unknowns could be reduced via instrumental constraints, by relating the baseline correction coefficients through a shared physical model. The $\sigma$ and $\mu$ fits directly determine baseline-level corrections of the $\mathbf{\bar{M}}^{-1}$ form introduced with \cref{eq:baseline_visibility_mixing_matrix}. Enforcing an antenna-level description would require this operator to admit an $\mathbf M^{-1}$ representation: its weight from input baseline $(k,l)$ to output baseline $(i,j)$ would then be a product $(\mathbf M^{-1})_{ik}(\mathbf M^{-1})_{jl}^{*}$. Constructing these weights from a smaller set of shared antenna-level coefficients would reduce the number of independent parameters. Array symmetry could also reduce the number of independent coefficients. For example, consider one baseline being used to correct another. If their mirror images elsewhere in the array have equivalent antenna surroundings and feed orientations, the two corrections could share the same underlying instrumental coupling parameters.

One could additionally constrain the coefficients across the frequency axis by representing their complex amplitudes with a polynomial or another suitable basis. The independent per-channel values could thus be replaced by a smaller set of parameters, just enough to retain the spectral structure associated with the longest coupling delays.

If the coupling mechanism is stable over the observing interval, successive integrations can provide further constraints as Earth rotation changes the sky response. Describing those integrations jointly as different views of the same sky observed via the same instrumental response, rather than treating every isolated visibility as an independent unknown, could help distinguish the sky's changing contribution from the effect of the shared instrumental parameters. We emphasize, however, that the optimal $\mu$ coefficients can depend on the illumination even when the coupling mechanism is fixed, as illustrated by the two distinct fields in \cref{fig:visibility_coefficient_field_transfer}. Nonetheless, the strong suppression obtained with one shared coefficient set over 2-hour observing intervals shows that coefficients need not be determined on a per-LST basis to outperform the sky-agnostic fits. Such assumptions about the stability of coupling mechanisms over time have been explored in \citet{kern_et_al_2019, kern_et_al_2020, rath_et_al_2025} to develop fringe-rate and delay filters based on coupling's characteristic imprint on visibilities.

Spectral redundancy offers another possible source of sky constraints \citep{cox_et_al_2024}. Baselines and frequencies with similar $\vec u=\vec b\nu/c$ probe similar foreground Fourier modes when the isolated beam and foreground emission evolve sufficiently slowly across the band. Their smooth evolution could relate the unknown uncoupled visibilities, while a coupling model accounts for the additional chromatic structure in the measured responses. We note that this same spectrally redundant formalism could also supply additional correction templates from neighboring channels. This would expand the model's flexibility rather than provide an additional constraint. We leave this extension to future work. Unlike the fractional-baseline test in \cref{subsec:kernel_template_enrichment}, it would require treating the frequency dependence of the coupled beams and sky explicitly: neighboring-channel visibilities cannot generally be replaced by shifted copies of a single-frequency response.

\subsection[Requirements for practical correction]{Requirements for practical correction}
\label{subsec:requirements_for_practical_correction}

Applying these ideas to real arrays will also require relaxing some of our instrumental assumptions. Using the same embedded beam for every element neglects the different neighborhoods of antennas in a real array. In particular, an edge response is not generally obtained by simply deleting some coupling coefficients from the central element's response, so an element-dependent correction may require both different weights and additional templates. Noise, flagging, and calibration will also affect the estimation and application of the correction; self-coupling, for example, is degenerate with an antenna gain.

Preserving the cosmological signal is a further requirement. Our $\mu$ fits minimize the error relative to foreground-only isolated visibilities, and may exploit cancellations specific to that foreground sky. The same correction could attenuate or distort an added 21~cm component. If its coefficients are estimated from observations, the fit could also absorb cosmological emission or foreground-model errors into the inferred coupling. Strong foreground suppression therefore does not establish signal preservation. Signal-injection and recovery tests, including re-estimation of the coefficients, are needed to quantify this risk.

Within the idealized conditions studied here, the fits demonstrate that linear visibility corrections can remove much of the simulated coupling despite the small antenna-mode fraction in the beams. Achieving comparable suppression in observations requires estimating those corrections with the information actually available, while demonstrating that the cosmological signal is preserved.
\section{Conclusion}
\label{sec:conclusion}

We have extended holographic back-projection to localize mutual-coupling perturbations around neighboring array elements, making their spatial structure visible. We established a coupling mode decomposition---inspired by antenna scattering theory, and expanded upon with a definition of our own---that explains the shortcomings of some linear combination coupling models. Applied to the 37-element HERA-like dish array, this decomposition shows that translated isolated responses capture a band-median fraction of only about $1\%$ of the beam perturbation power over $60$--$220$~MHz, a fraction that increases when the dishes are removed and the feeds are pointed toward the sky.

This small antenna-mode fraction does not translate directly into a poor visibility correction, because forming beam products changes the relative weighting of the coupling modes, while sky convolution and baseline sampling may further affect which perturbations dominate. Allowing independent kernel weights additionally relaxes the relationships imposed by the antenna-mode coupling model, enabling the correction to remove part of the structural coupling as well. In our noiseless, perfectly redundant dish-array simulation, the projected antenna-mode inverse removes roughly two thirds of the coupling power, and fitting the coefficients directly to the visibilities achieves deeper suppression (\cref{subsec:beam_and_visibility_fitted_corrections}). The optimal visibility-space coefficients also depend on the illuminating sky, so the suppression achieved on one field does not automatically carry over to another.

For the chosen template set, sampled visibilities, and least-squares objective, the $\mu$ fit provides an idealized benchmark: it finds the smallest residual available within that model when the isolated visibilities are known. This is not a bound on all possible linear corrections, nor a guarantee of recovery for an unknown sky. It does show that the small antenna-mode fraction in the beams does not preclude substantial coupling suppression by linear combinations of visibilities. The remaining challenge is to retain that suppression with observationally justified constraints, element-dependent responses, noise, and flagging, while demonstrating preservation of the cosmological signal.

\section*{Acknowledgements}

We thank James Aguirre, Kai-Feng Chen, Daniel Jacobs, and Miguel Morales for insightful discussions that helped shape this paper. 
This material is based upon work supported by the National Science Foundation under Grant Nos. 1636646 and 1836019 and institutional support from the HERA collaboration partners. This research is funded in part by the Gordon and Betty Moore Foundation through Grant GBMF5212 to the Massachusetts Institute of Technology.

VM and JNH acknowledge support from the MIT School of Science and the Gordon and Betty Moore Foundation (the latter through grant GBMF5212 to the Massachusetts Institute of Technology). JSD, AP, and TC acknowledge support from the U.S.\ National Science Foundation under Grant No.~2606478. RP acknowledges support from the Faculty of Arts \& Science at the University of Toronto and the Dunlap Institute. The Dunlap Institute is funded through an endowment established by the David Dunlap family and the University of Toronto. NK acknowledges support from the National Aeronautics and Space Administration (NASA) through the NASA Hubble Fellowship grant \#HST-HF2-51533.001-A awarded by the Space Telescope Science Institute, which is operated by the Association of Universities for Research in Astronomy, Incorporated, under NASA contract NAS5-26555. EdLA acknowledges support from UKRI STFC via an STFC Ernest Rutherford Fellowship (ST/V004425/1). The simulations presented here use a simplified version of the HERA feed model developed and validated by NF and EdLA.
\appendix
\crefalias{section}{appendix}

\section{Electromagnetic simulations}
\label{app:cst_setup}

\begin{figure*}[h!]
  \centering
  \setlength{\tabcolsep}{2pt}
  \begin{tabular}{ccc}
  \includegraphics[width=0.32\textwidth]{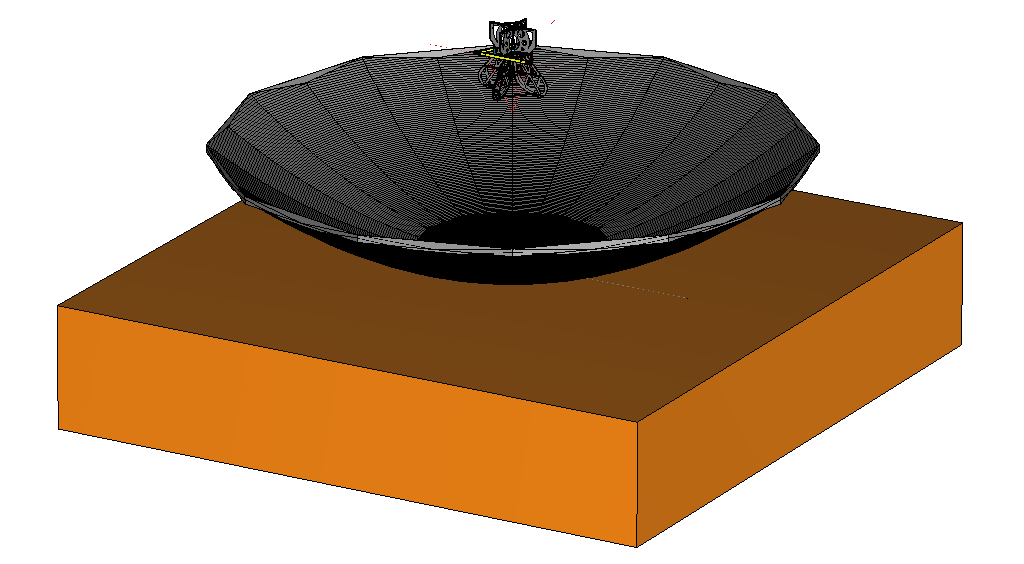} &
  \includegraphics[width=0.32\textwidth]{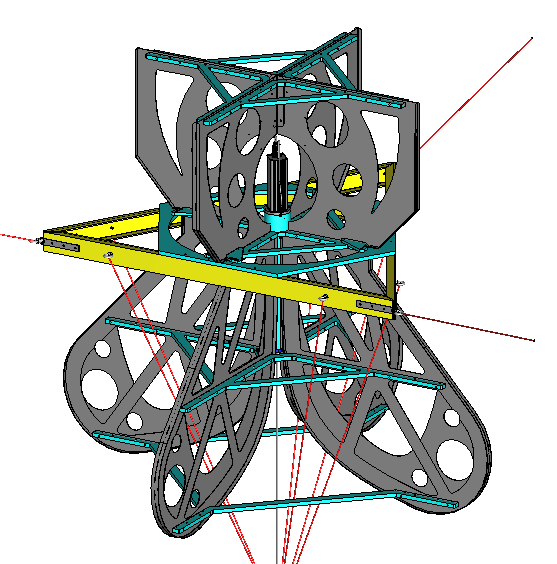} &
  \includegraphics[width=0.32\textwidth]{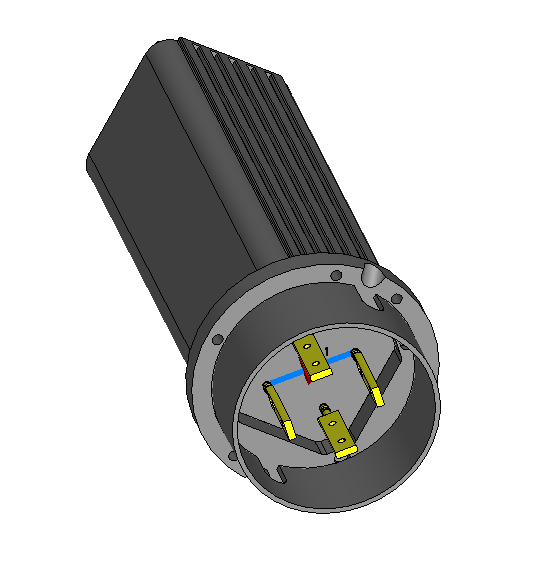} \\
  \includegraphics[width=0.32\textwidth]{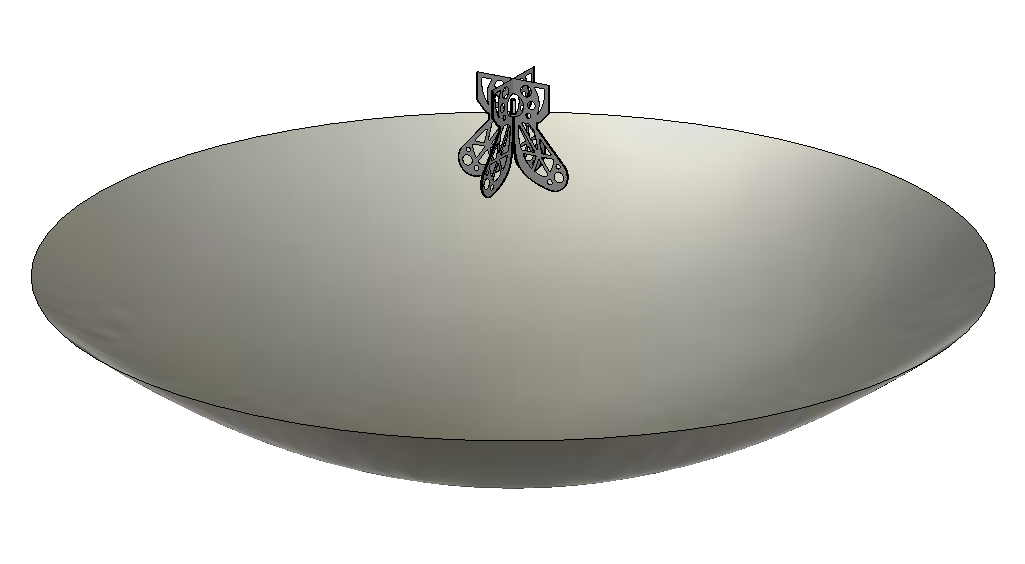} &
  \includegraphics[width=0.32\textwidth]{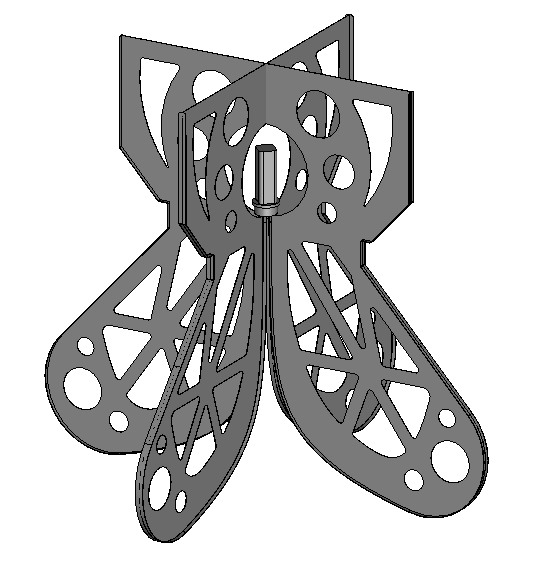} &
  \includegraphics[width=0.32\textwidth]{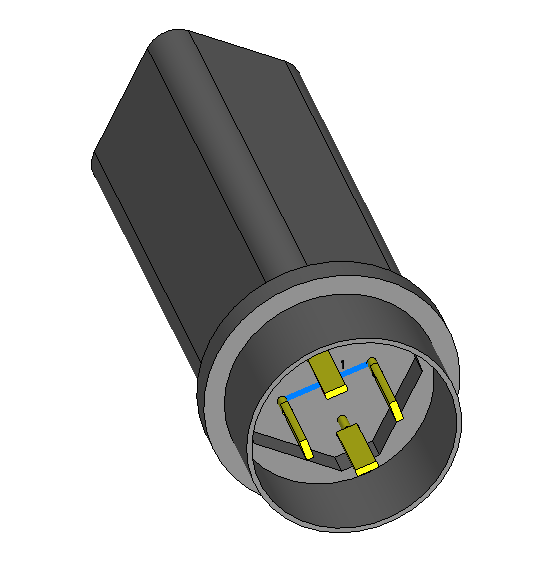}
  \end{tabular}
  \caption{Comparison of the detailed \citet{fagnoni_et_al_2021} model (top) and the simplified model used here (bottom). Left: full dish. Center: feed. Right: front-end (RF box). The simplified model uses an analytic parabola with no soil or support structures.\label{fig:detailed_and_simplified_cst_models}}
\end{figure*}

The electromagnetic simulations use a CST Studio Suite model based on the detailed HERA element of \citet{fagnoni_et_al_2021}, simplified to reduce the mesh-cell count enough for large multi-element simulations: we remove all support structures and many small features, replace the dish with an analytic parabola, and omit the soil. \Cref{fig:detailed_and_simplified_cst_models} compares the detailed and simplified models. These simplifications reduce fidelity to the existing HERA dishes while retaining the overall geometry expected to drive the dominant coupling features, providing a realistic case study of strong coupling without attempting to reproduce the instrument's exact response.

The simulations were run in CST with the time-domain solver over the $60$--$220$~MHz band at 1~MHz frequency resolution. Each element is driven at a single feed port while the remaining ports are terminated in the load $Z_L$ appropriate to the run: the reference impedance $Z_0 = 100~\Omega$ unless otherwise indicated, and the special loads of \cref{tab:antenna_mode_reference_loads} (impedance match, short, and open) for the mode-separation runs of \cref{app:antenna_mode_load_conventions}, with the conjugate match realized as a lumped element at $150$~MHz only. For each configuration, the solver outputs the complex far-field voltage pattern $\vec E(\theta,\phi)$ of the central element and the full scattering matrix $S$, from which the analytic antenna-mode coefficients of \cref{eq:load_dependent_coupling_from_s_parameters} are evaluated. We simulate an isolated element, \texttt{lin2} at the $14.6$~m HERA spacing under each load convention, and the \texttt{hex37} array of \cref{fig:hex37_layout_and_coupled_aperture}. The feed-only family used throughout---an isolated feed, \texttt{lin2}, and \texttt{hex37} in the same layouts---uses the same feeds with the dishes removed and the feeds pointed toward the sky, with all ports terminated at $Z_0$.

\section{From visibilities to beam kernels}
\label{app:beam_kernel_derivation}

We derive the convolution form of
\cref{eq:visibility_sky_kernel_convolution} and the aperture cross-correlation of
\cref{eq:aperture_cross_correlation_kernel}.
We use the coplanar geometry introduced in
\cref{subsec:beam_kernel_definition}, with $b_z=w=0$ and
$\vec u=(b_x,b_y)/\lambda$.
On the front hemisphere,
$d^2\hat s=d\ell\,dm/n$, where
$n=\sqrt{1-\ell^2-m^2}$, so the visibility integral is
\begin{equation}
  V_{ij}(\vec u)
  =\iint_{\ell^2+m^2\leq1}
  \frac{T(\ell,m)}{n}\,
  \vec E_i(\ell,m)\cdot\vec E_j^*(\ell,m)\,
  e^{-2\pi i(u\ell+vm)}\,d\ell\,dm.
  \label{eq:visibility_integral_in_direction_cosines}
\end{equation}
With the Fourier convention
$\mathcal F[f](\vec u)
=\iint f(\ell,m)e^{-2\pi i(u\ell+vm)}\,d\ell\,dm$,
the inverse transform integrates over $du\,dv$ with the opposite sign in the exponent and no prefactor.
All functions are extended by zero outside the visible disk.
The convolution theorem therefore gives
\begin{equation}
  V_{ij}(\vec u)
  =\left(
  \mathcal F[T/n]*
  \mathcal F[\vec E_i\cdot\vec E_j^*]
  \right)(\vec u)
  =\bigl(\tilde T*K_{ij}\bigr)(\vec u).
\end{equation}
The solid-angle Jacobian is included in the sky factor
$\tilde T=\mathcal F[T/n]$, while the beam kernel is
$K_{ij}=\mathcal F[\vec E_i\cdot\vec E_j^*]$.
The convolution integrates over dimensionless $uv$ coordinates, so no wavelength prefactor is needed.

The aperture fields of \cref{eq:beam_to_aperture_transform}
use physical coordinates:
$\vec A_i(\vec x)=\mathcal F[\vec E_i](\vec x/\lambda)$.
Applying the correlation theorem and changing variables from
$\vec u\,'$ to $\vec x=\lambda\vec u\,'$ gives
\begin{equation}
\begin{aligned}
  K_{ij}(\vec u)
  &=\int
  \mathcal F[\vec E_i](\vec u\,')\cdot
  \mathcal F[\vec E_j]^*(\vec u\,'-\vec u)\,d^2u'\\
  &=\frac{1}{\lambda^2}\int
  \vec A_i(\vec x)\cdot
  \vec A_j^*(\vec x-\lambda\vec u)\,d^2x.
\end{aligned}
\end{equation}
The factor $\lambda^{-2}$ follows from
$d^2u'=d^2x/\lambda^2$ and is retained explicitly.
The peak normalization used in the figures is a separate display convention.
\section{All-paths coupling coefficients}
\label{app:all_paths_coupling_coefficients}

Following \citet{josaitis_et_al_2022}, a wave received at element $k$ and re-radiated toward element $i$ contributes a voltage $\epsilon^{(1)}_{ik}v_k^{(0)}$ to $i$'s coupled output, where $v_k^{(0)}$ is the direct (uncoupled) voltage at $k$ and $\epsilon^{(1)}_{ik} \in \mathbb{C}$ is the single-bounce coupling coefficient between elements $i$ and $k$. It is a function of the antenna patterns, the load impedance, and the propagator (an amplitude-decay and phase-modulating prefactor that depends on the distance between the two elements involved). A first-order treatment that includes only single-bounce paths writes
\begin{equation}
  v_i^{(1)} = v_i^{(0)} + \sum_{k \neq i} \epsilon^{(1)}_{ik} v_k^{(0)},
  \label{eq:voltage_single_scattering_order}
\end{equation}
and is what \citet{josaitis_et_al_2022} and \citet{rath_et_al_2025} adopt. Going beyond first order means feeding coupled voltages back into \cref{eq:voltage_single_scattering_order}. The iteration raises the superscript of the voltages \emph{inside} the sum, while retaining the direct contribution $v_i^{(0)}$ outside it:
\begin{equation}
  v_i^{(2)} = v_i^{(0)} + \sum_{k \neq i} \epsilon^{(1)}_{ik} v_k^{(1)}
            = v_i^{(0)} + \sum_{k \neq i} \epsilon^{(1)}_{ik}\Big(v_k^{(0)} + \sum_{l \neq k}\epsilon^{(1)}_{kl} v_l^{(0)}\Big),
  \label{eq:voltage_through_second_scattering_order}
\end{equation}
and one more iteration gives
\begin{equation}
  v_i^{(3)} = v_i^{(0)} + \sum_{k \neq i} \epsilon^{(1)}_{ik}\Big(v_k^{(0)} + \sum_{l \neq k}\epsilon^{(1)}_{kl}\big(v_l^{(0)} + \sum_{m \neq l}\epsilon^{(1)}_{lm} v_m^{(0)}\big)\Big).
  \label{eq:voltage_through_third_scattering_order}
\end{equation}

Indexed this way---by the number of reflection events---the bookkeeping quickly becomes unwieldy: each order nests another sum, and the number of terms grows exponentially. But every term, fully expanded, is a product of single-bounce coefficients along a \emph{path} that starts at some element and ends at $i$, multiplying the starting element's direct voltage. This suggests reorganizing the expansion by \emph{endpoints}: collect, for each pair $(i,k)$, every path from $k$ to $i$ regardless of its length. We set $\epsilon^{(1)}_{ii}=0$, so the unrestricted sums below exclude immediate self-hops. Expanding \cref{eq:voltage_through_third_scattering_order} and regrouping terms by the element whose direct voltage they carry,
\begin{equation}
  v_i^{(3)} = v_i^{(0)} + \sum_{k=1}^{N}\Big(\epsilon^{(1)}_{ik} + \sum_l\epsilon^{(1)}_{il}\epsilon^{(1)}_{lk} + \sum_{l,m}\epsilon^{(1)}_{im}\epsilon^{(1)}_{ml}\epsilon^{(1)}_{lk}\Big) v_k^{(0)}:
  \label{eq:third_order_voltage_grouped_by_endpoints}
\end{equation}
the nested structure has disappeared into the bracket, which collects the paths with the same endpoints. Provided successive reflections become sufficiently weak for their sum to converge, taking the number of reflections to infinity turns the bracket into the all-paths coefficient
\begin{equation}
  \epsilon_{ik} \equiv \epsilon^{(1)}_{ik} + \sum_l \epsilon^{(1)}_{il}\epsilon^{(1)}_{lk} + \sum_{l,m}\epsilon^{(1)}_{im}\epsilon^{(1)}_{ml}\epsilon^{(1)}_{lk} + \dots,
  \label{eq:all_paths_coupling_coefficient_matrix}
\end{equation}
the sum running over all paths from $k$ to $i$ of every length. At each frequency, $\epsilon_{ik}$ is a single complex number that absorbs every multi-bounce path from $k$ to $i$, and the coupled voltage becomes
\begin{equation}
  v_i^{\rm c} = v_i^{(0)} + \sum_{k=1}^{N} \epsilon_{ik} v_k^{(0)},
  \label{eq:all_paths_voltage_mixing}
\end{equation}
of the same form as \cref{eq:voltage_single_scattering_order}, now including the self-response correction and all orders of reflection within the linear re-radiation model. Rather than enumerating the individual paths, we fit these all-paths coefficients directly in aperture space. Thus, every $\epsilon_{ik}$ in the main text includes any number of ``bounces'' between elements, with the superscript $(1)$ reserved for the single-bounce coefficient used only in this appendix.

\section{Load conventions for the antenna mode}
\label{app:antenna_mode_load_conventions}

This appendix briefly reviews the four load conventions used in the scattering literature to define the antenna-mode coefficient $\epsilon(Z_L)$ of \cref{eq:load_dependent_scattered_field_decomposition}, and the corresponding special load $Z_L'$, and compares the decompositions they produce with the projected definition adopted in \cref{subsec:projected_antenna_mode_definition}.

The first two conventions build the special load from the impedance inherent to the antenna: \citet{knott_et_al_2004} use $Z_L'=Z_A$ (the antenna impedance) for its ``intuitive appeal,'' while \citet{green_1963} uses the conjugate match $Z_L'=Z_A^*$, corresponding to maximum power transfer. Both produce re-radiation proportional to $\vec E_{\rm iso}$; in fact, neglecting ohmic losses, the re-radiated power equals the absorbed power whenever $\Re(Z_L)=\Re(Z_A)$, which is true for both of those definitions. 

The two remaining conventions instead fix the load at an extreme. \citet{hansen_1989} adopts the short circuit $Z_L'=0$---a choice offered without explicit justification, but sufficiently widespread to warrant inclusion here---whereas \citet{collin_2003} and \citet{best_kaanta_2009} favor the open circuit $Z_L'=\infty$. The latter aligns naturally with the minimum-scattering-antenna (MSA) framework of \citet{kahn_kurss_1965}: an MSA is an idealized antenna which, by definition, does not admit structural-mode scattering. It has the property that it becomes electrically invisible when open-circuited, so any scattering remaining at $Z_L=\infty$ is attributed to structure beyond the MSA.

For each convention, the antenna-mode field is proportional to the difference between the terminal current at the operating load $Z_L$ and that at the special load $Z_L'$. With the same excitation, the receiving-antenna circuit gives $I(Z_L)=V_{\rm oc}/(Z_A+Z_L)$, where $V_{\rm oc}$ is the open-circuit voltage, so that
\begin{equation*}
  I(Z_L)-I(Z_L')=V_{\rm oc}\left(\frac{1}{Z_A+Z_L}-\frac{1}{Z_A+Z_L'}\right).
\end{equation*}
For the impedance and conjugate matches, this difference can also be written using
\begin{equation}
  \Gamma_{\!A}(Z_L) = \frac{Z_L-Z_A}{Z_L+Z_A},\quad\tilde{\Gamma}_{\!A}(Z_L) = \frac{Z_L-Z_A^*}{Z_L+Z_A}.
\end{equation}
These current differences describe the load dependence; to obtain a dimensionless beam coefficient, we use the two-element (four-port) scattering parameters. Here the operating load is fixed at $Z_L=Z_0=100~\Omega$, while $Z_L'$ varies between conventions. Let $a_1$ be the incident wave driving the central feed, $b_3$ the wave leaving the network toward the neighbor's load, and $a_3$ the wave reflected back into the network. With the other ports matched to $Z_0$,
\begin{equation*}
  b_3=S_{31}a_1+S_{33}a_3,\qquad
  a_3=\Gamma_0(Z_L')b_3,\qquad
  \Gamma_0(Z)=\frac{Z-Z_0}{Z+Z_0}.
\end{equation*}
The operating load reflects no wave back into port 3. Subtracting the special-load beam from that operating-load beam therefore gives a contribution with coefficient $-a_3/a_1$ multiplying the neighbor's embedded pattern, normalized to the same incident wave. Approximating this pattern by the translated isolated response gives
\begin{equation}
  \epsilon(Z_0) \simeq -\frac{a_3}{a_1}
  = -\,\frac{\Gamma_0(Z_L')\,S_{31}}{1-\Gamma_0(Z_L')\,S_{33}},
  \label{eq:load_dependent_coupling_from_s_parameters}
\end{equation}
where the denominator accounts for repeated reflections between the neighbor's load and the reference network. The coefficient is evaluated at the operating load $Z_0$, but depends on which special load defines the structural mode. For the symmetric \texttt{lin2} system, we use the input impedance inferred from $S_{11}$, with the other ports terminated at $Z_0$, to set $Z_A$. Since $Z_0$ is real, the four conventions give $\Gamma_0(Z_L')\in\{S_{11},\,S_{11}^*,\,-1,\,+1\}$, respectively. \Cref{tab:antenna_mode_reference_loads} summarizes these reflection coefficients and the corresponding current differences.

\begin{table}[h!]\caption{Four conventions for the special load impedance $Z_L'$. The current difference sets the load dependence of the antenna-mode field; $\Gamma_0(Z_L')$ is used in the beam-coefficient estimate of \cref{eq:load_dependent_coupling_from_s_parameters}.}
\centering
  \begin{tabular}{lllll}
  \hline\hline
  Name & $Z_L'$ & $I(Z_L)-I(Z_L')$ & $\Gamma_0(Z_L')$ & Reference\\
  \hline
Imped. match & $Z_A$    & $-\Gamma_{\!A}(Z_L)I(Z_A)$ & $S_{11}$ & \citet{knott_et_al_2004}\\
Conj. match & $Z_A^*$   & $-\tilde{\Gamma}_{\!A}(Z_L)I(Z_A^*)$ & $S_{11}^*$ & \citet{green_1963}\\
Short circuit   & $0$      & $I(Z_L)-I(0)$ & $-1$ & \citet{hansen_1989}\\
Open circuit   & $\infty$      & $I(Z_L)$ & $+1$ & \citet{collin_2003}, \citet{best_kaanta_2009}\\
\hline\hline
\end{tabular}\label{tab:antenna_mode_reference_loads}
\end{table}



\begin{figure}[h!]
  \centering
  \includegraphics[width=0.85\columnwidth]{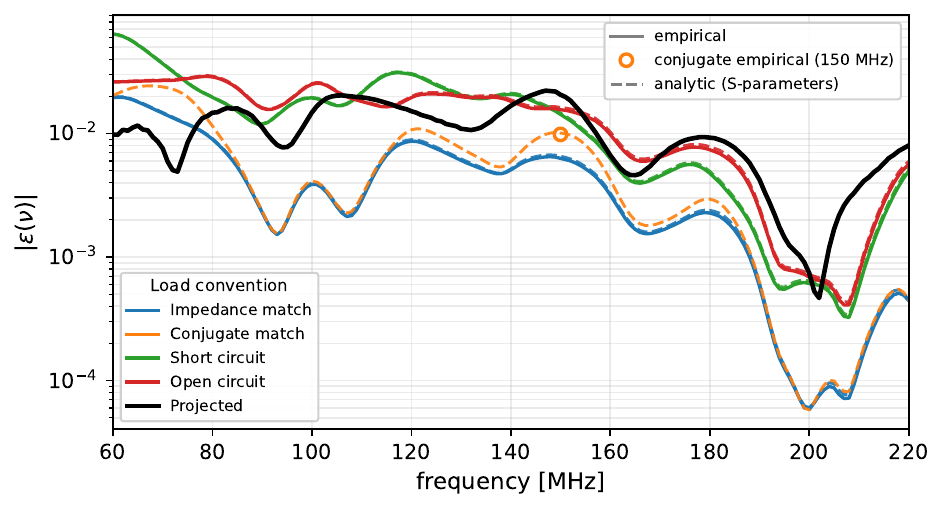}
  \caption{Coupling coefficient $|\epsilon(\nu)|$ for all five definitions across the HERA band, at the operating load $Z_0=100~\Omega$. The four load conventions of \cref{tab:antenna_mode_reference_loads} are shown as analytic estimates (\cref{eq:load_dependent_coupling_from_s_parameters}, dashed) and empirically (solid, same color per load); the empirical curves come from differencing the reference ($Z_0$) coupled beam against each special-load beam and projecting onto the shifted isolated pattern. The conjugate-match empirical result is available only at 150~MHz and is shown by an open circle. The projected coefficient $|\epsilon_{\rm proj}|$ of \cref{eq:single_neighbor_projection_coefficient} is shown in black.\label{fig:load_and_projected_coupling_coefficients}}
\end{figure}

\Cref{fig:load_and_projected_coupling_coefficients} compares all five coefficient definitions across the band. The analytic curves use \cref{eq:load_dependent_coupling_from_s_parameters} with the special-load reflection coefficients in \cref{tab:antenna_mode_reference_loads}. To obtain each empirical coefficient, we simulate the neighbor at the special load $Z_L'$ that extinguishes its antenna mode, subtract that beam from the reference coupled beam at $Z_0=100~\Omega$, and project the difference onto the shifted isolated pattern. This projection measures the overlap with the isolated template, whereas the analytic estimate treats the neighbor's embedded pattern as that template. The projected definition $\epsilon_{\rm proj}$ instead uses the full perturbation $\vec A_{\rm tot}-\vec A_{\rm iso}$, without a special-load simulation. The conjugate match requires separate treatment: $Z_A^*(\nu)$ is anti-causal and cannot be realized as a passive broadband load in the simulator. We evaluate its empirical coefficient only at 150~MHz, where the load can be realized as a lumped R+C or R+L element.


\FloatBarrier
\section[Maps of the coupling modes]{Maps of the coupling modes}
\label{app:supplementary_coupling_maps}

We supplement the visibility-space comparisons of \cref{sec:coupling_in_simulated_visibilities} with simulated maps of the coupling modes for the fiducial field and Fornax~A transit.

\begin{figure}[h!]
  \centering
  \includegraphics[width=\textwidth]{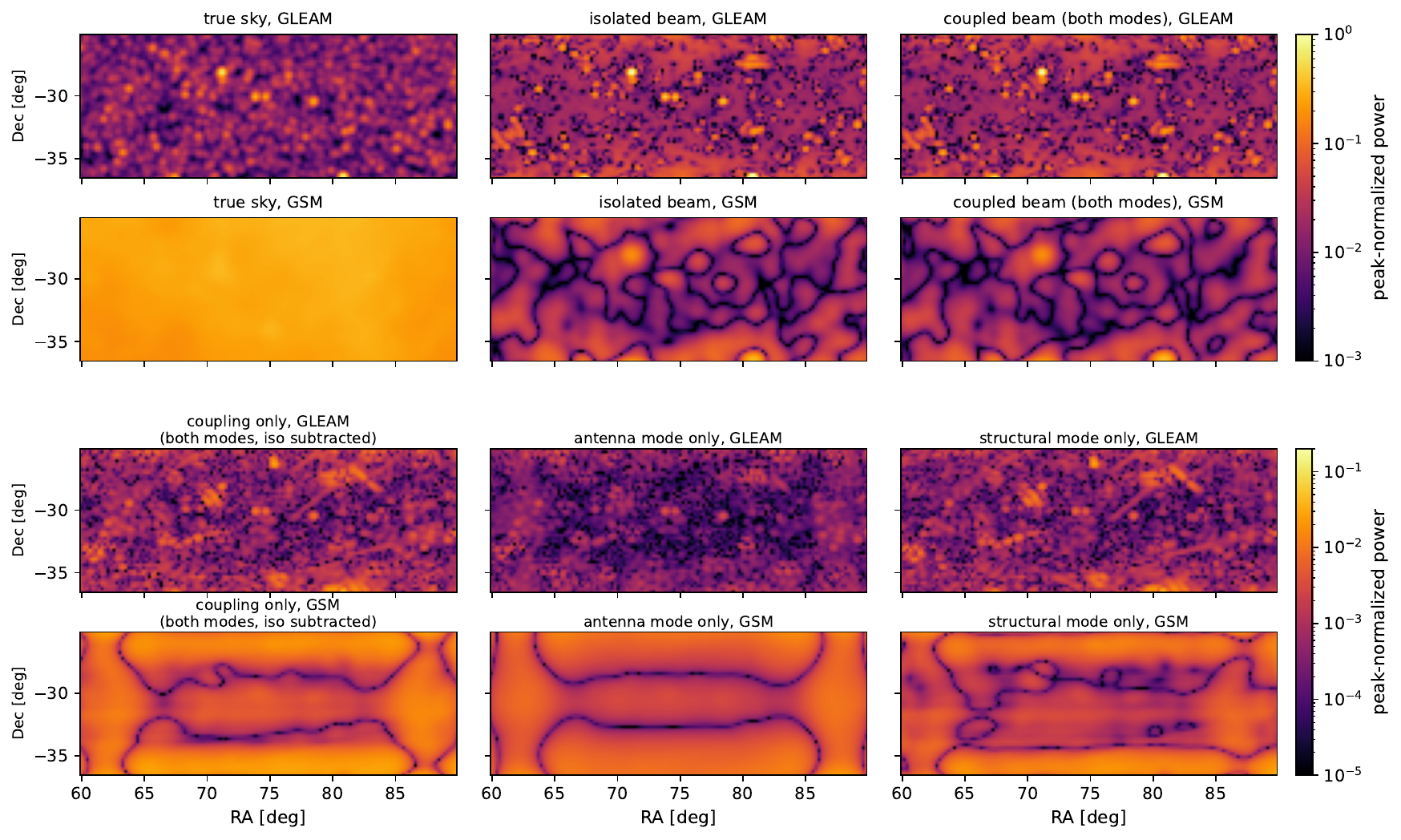}
  \caption{Maps of the fiducial field at 150\,MHz for a point-source sky (GLEAM, rows 1 and 3) and a diffuse sky (GSM, rows 2 and 4). Top group shows the true sky (column 1), isolated beam map (column 2), and coupled beam map (column 3). Bottom group shows perturbations due to both coupling modes (column 1), antenna mode only (column 2), and structural mode only (column 3). All panels are peak-normalized (the true-sky panels to the true-sky peak; the maps to the isolated map's peak). No strong characteristic feature associated with either coupling mode is visible, beyond broad, faint stripes (bottom row). See text for the simulation and mapping details.}\label{fig:coupling_mode_sky_images}
\end{figure}

\begin{figure}[h!]
  \centering
  \includegraphics[width=\textwidth]{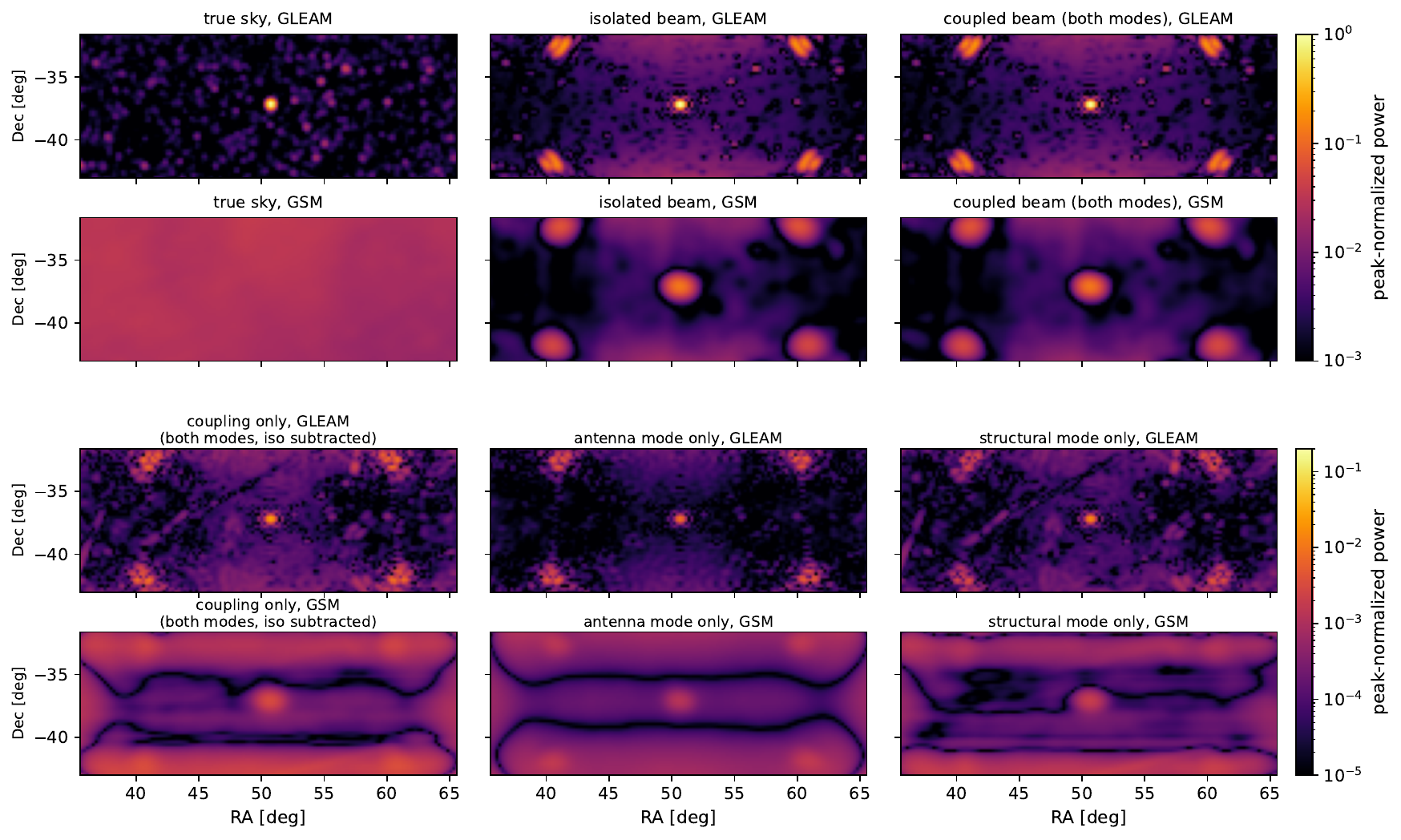}
  \caption{Maps of the Fornax~A transit at 150\,MHz, with the same panel arrangement and normalization convention as \cref{fig:coupling_mode_sky_images}. The point-source and diffuse skies are shown separately, and the normalization uses the peaks of this field's true sky and isolated maps.}
  \label{fig:fornax_coupling_mode_sky_images}
\end{figure}

\Cref{fig:coupling_mode_sky_images,fig:fornax_coupling_mode_sky_images} show maps of the fiducial field and Fornax~A transit produced from the simulated visibilities via direct optimal mapping \citep{xu_et_al_2022} at 150\,MHz, with the two skies (GLEAM, GSM) shown separately. To enhance resolution, we extend the array to a larger offset-diamond layout ($300$ elements, $278$~m maximum baseline) whose shortest spacings reproduce the \texttt{hex37} lattice; despite this larger layout, we maintain the identical-element approximation for the primary beams, i.e., each element is simulated with the beam obtained from the full-wave simulation of \texttt{hex37}'s central element. Each pixel of a $30^\circ\times15^\circ$ facet accumulates the fringe- and beam-weighted visibilities over all baselines and integrations, and is then normalized by its accumulated squared beam, so the displayed maps have uniform gain across the well-measured region; the axes are cropped to that region.

We note that map space does not show bright distinctive features of coupling, although this could be due to our limited electromagnetic model. In both modes, the drift accumulation averages away the east--west fringe components, leaving faint horizontal bands whose brightness varies in the north--south direction (bottom row of \Cref{fig:coupling_mode_sky_images,fig:fornax_coupling_mode_sky_images}). The antenna mode's fringes are broader, likely a consequence of the limited extent of its kernel perturbation in baseline space; conversely, the structural mode's perturbations yield finer striping and a sharper response to point sources.

\FloatBarrier
\setcounter{biburlnumpenalty}{100}
\begingroup
\setlength{\emergencystretch}{1em}
\printbibliography
\endgroup

\end{document}